\documentclass[aps,prb,superscriptaddress,twocolumn,floatfix]{revtex4}
\usepackage{amsmath}
\usepackage{graphicx}

\begin{document}

\title{The {\sc Siesta} method for ab initio order-$N$ materials simulation}

\author{ Jos\'e M. Soler }
\affiliation{ Dep.\ de F\'{\i}sica de la Materia Condensada, C-III,
              Universidad Aut\'{o}noma de Madrid,
              E-28049 Madrid, Spain }

\author{ Emilio Artacho }
\affiliation{ Department of Earth Sciences, 
              University of Cambridge,
              Downing St., Cambridge CB2 3EQ, United Kingdom }

\author{ Julian D. Gale }
\affiliation{ Department of Chemistry, 
              Imperial College of Science, Technology and Medicine,
              South Kensington SW7 2AY, United Kingdom }

\author{ Alberto Garc\'{\i}a }
\affiliation{ Departamento de F\'{\i}sica de la Materia Condensada,
              Universidad del Pa\'{\i}s Vasco,
              Apt. 644, 48080 Bilbao, Spain }

\author{ Javier Junquera }
\affiliation{ Dep.\ de F\'{\i}sica de la Materia Condensada, C-III,
              Universidad Aut\'{o}noma de Madrid,
              E-28049 Madrid, Spain }
\affiliation{ Institut de Physique, B\^atiment B5, Universit\'e de Li\`ege,
              B-4000 Sart-Tilman, Belgium }

\author{ Pablo Ordej\'on }
\affiliation{ Institut de Ci\`encia de Materials de Barcelona, CSIC,
              Campus de la UAB, Bellaterra,
              08193 Barcelona, Spain }

\author{ Daniel S\'anchez-Portal }
\affiliation{ Dep. de F\'{\i}sica de Materiales and DIPC,
              Facultad de Qu\'{\i}mica, UPV/EHU,
              Apt. 1072, 20080 Donostia, Spain }

\date{\today}

\begin{abstract}
   We have developed and implemented a self-consistent density functional 
method using standard norm-conserving pseudopotentials and a flexible, 
numerical LCAO basis set, which includes multiple-zeta and polarization 
orbitals.
   Exchange and correlation are treated with the local spin density
or generalized gradient approximations.
   The basis functions and the electron density are projected on
a real-space grid, in order to calculate the Hartree and 
exchange-correlation  potentials and matrix elements, with a number
of operations that scales linearly with the size of the system.
   We use a modified energy functional, whose minimization
produces orthogonal wavefunctions and the same energy and density 
as the Kohn-Sham energy functional, without the need of an
explicit orthogonalization.
   Additionally, using localized Wannier-like electron wavefunctions  
allows the computation time and memory, required to minimize 
the energy, to also scale linearly with the size of the system.
   Forces and stresses are also calculated efficiently and accurately, 
thus allowing structural relaxation and molecular dynamics simulations.
\end{abstract}

\pacs{PACS numbers: 71.15.Mb, 71.15.Nc}


\maketitle

\section{Introduction}

   As the improvements in computer hardware and software allow the
simulation of molecules and materials with an increasing number of 
atoms $N$, the use of so-called order-$N$ algorithms, in which the 
computer time and memory scales linearly with the simulated system 
size, becomes increasingly important.
   These ${\cal O}(N)$ methods were developed during the 1970's and 80's
for long range forces 
 \cite{Greengard1994}
and empirical interatomic potentials
 \cite{Hockney-Eastwood1988}
but only in the last 5-10
years for the much more complex quantum-mechanical methods,
in which atomic forces are obtained by solving the interaction 
of ions and electrons together
 \cite{Car-Parrinello1985}.
   Even among quantum mechanical methods, there are very different 
levels of approximation: empirical or semiempirical orthogonal
tight-binding methods are the simplest ones
 \cite{Goringe-Bowler-Hernandez1997:revTB,Ordejon1998:revTBON}; 
`ab-initio' nonorthogonal-tight-binding 
and nonself-consistent 
Harris-functional methods are next
 \cite{Ordejon1993,Sankey-Niklewski1989}; 
and fully self-consistent density 
functional theory (DFT) methods are the most complex and reliable
 \cite{Payne1992:RMP}.
   The implementation of ${\cal O}(N)$ methods in quantum-mechanical
simulations has also followed these steps, with several methods
already well established within the tight-binding formalism
 \cite{Ordejon1998:revTBON},
but much less so in self-consistent DFT
 \cite{Goedecker1999:RMP}.
   The latter also require, in addition to solving Schr\"odinger's equation,
the determination of the self-consistent Hamiltonian in ${\cal O}(N)$ iterations.
   While this is difficult using plane waves, a localized basis set
appears to be the natural choice.
   One proposed approach are the `blips' of Hernandez and Gillan
 \cite{Hernandez-Gillan1995},
regularly-spaced Gaussian-like splines that can be systematically
increased, in the spirit of finite-element methods, although at a
considerable computational cost.

   We have developed a fully self-consistent DFT, 
based on a flexible linear combination of atomic orbitals (LCAO)
basis set, with essentially perfect ${\cal O}(N)$ scaling.
   It allows extremely fast simulations using minimal basis sets and very 
accurate calculations with complete multiple-zeta and polarized bases,
depending on the required accuracy and available computational power.
   In previous papers
\cite{Ordejon1996,SanchezPortal1997}
we have described preliminary versions of this method, 
that we call {\sc Siesta} (Spanish Initiative for Electronic Simulations
with Thousands of Atoms).
   There is also a review 
\cite{Ordejon2000:revSIESTAappl}
of the tens of studies performed with it, in a wide variety of systems,
like metallic surfaces, nanotubes, and biomolecules.
   In this work we present a more complete description of the method,
as well as some important improvements.

   Apart from that of Born and Oppenheimer, the most basic approximations 
concern the treatment of exchange and correlation, and the use of
pseudopotentials.
   Exchange and correlation (XC) are treated within Kohn-Sham DFT
 \cite{Kohn-Sham1965}.
   We allow for both the local (spin) density approximation 
 \cite{Perdew-Zunger1981} (LDA/LSD)
or the generalized gradient approximation  
 \cite{Perdew-Burke-Ernzerhof1996} (GGA).
   We use standard norm-conserving pseudopotentials
 \cite{Hamann-Schluter-Chiang1979,Bachelet1982}
in their fully non-local  form
 \cite{Kleinman-Bylander1982}.
   We also include scalar-relativistic effects and the nonlinear 
partial-core-correction to treat exchange and correlation in the core region
 \cite{Louie-Froyen-Cohen1982:NLCC}.

   The {\sc Siesta} code has been already tested and applied to dozens 
of systems and a variety of properties 
\cite{Ordejon2000:revSIESTAappl}.
   Therefore,  we will just illustrate here the convergence of a few
characteristic magnitudes of silicon, the architypical system of the 
field, with respect to the main precision parameters that characterize 
our method:
basis size (number of atomic basis orbitals);
basis range (radius of the basis orbitals);
fineness of the real-space integration grid; and
confinement radius of the Wannier-like electron states.
   Other parameters, like the $k$-sampling integration grid, are 
common to all similar methods and we will not discuss their
convergence here.

\section{Pseudopotential}

   Although the use of pseudopotentials is not strictly necessary with
atomic basis sets, we find them convenient to get rid of the core
electrons and, more importantly, to allow for the expansion of a
smooth (pseudo)charge density on a uniform spatial grid.
   The theory and usage of first principles norm-conserving 
pseudopotentials \cite{Hamann-Schluter-Chiang1979}
is already well established.
   {\sc Siesta} reads them in semilocal form
(a different radial potential $V_l(r)$ for each angular momentum $l$,
optionally generated scalar-relativistically
\cite{Kleinman1980,Bachelet-Schluter1982:relativistic})
from a data file that users can fill with their preferred choice.
   We generally use the Troullier-Martins parameterization
 \cite{Troullier-Martins1991}.
   We transform this semilocal form into the fully non-local form
proposed by Kleinman and Bylander (KB)
 \cite{Kleinman-Bylander1982}:
\begin{equation}
\label{VPS}
   \hat{V}^{PS} =  V_{local}(r) + \hat{V}^{KB}
\end{equation}
\begin{equation}
  \hat{V}^{KB} = \sum_{l=0}^{l^{KB}_{max}}
                 \sum_{m=-l}^l \sum_{n=1}^{N_l^{KB}}
         | \chi^{KB}_{lmn} \rangle v^{KB}_{ln}  \langle \chi^{KB}_{lmn} |
\label{VKB}
\end{equation}
\begin{equation}
\label{vKBln}
   v^{KB}_{ln} =  < \varphi_{ln} | \delta V_{l}(r) | \varphi_{ln} >
\end{equation}
where $\delta V_{l}(r) = V_{l}(r)-V_{local}(r)$.
$\chi^{KB}_{lmn}({\bf r}) = \chi^{KB}_{ln}(r) Y_{lm}(\hat{\bf r})$
(with $Y_{lm}(\hat{\bf r})$ a spherical harmonic)
are the KB projection functions
\begin{equation}
  \chi^{KB}_{ln}(r) =  \delta V_l(r)  \varphi_{ln}(r).
\label{xiKB}
\end{equation}
   The functions $\varphi_{ln}$ are obtained from the eigenstates
$\psi_{ln}$ of the semilocal pseudopotential (screened by the
pseudo-valence charge density) at energy $\epsilon_{ln}$
using the orthogonalization scheme proposed by
Bl\"ochl\cite{Blochl1990}:
\begin{equation}
  \varphi_{ln}(r)  =  \psi_{ln}(r) -  \sum_{n'=1}^{n-1} \varphi_{ln'}(r)
               \frac{ < \varphi_{ln'} | \delta V_l(r) | \psi_{ln} > }
                     { < \varphi_{ln'} | \delta V_l(r) | \varphi_{ln'} > }
\label{BlochlKB}
\end{equation}
\begin{eqnarray}
  [ -\frac{1}{2r} \frac{d^2}{d r^2} r
    & + & \frac{l(l+1)}{2r^2} + V_l(r) +  \nonumber \\
    & + & V^H(r) + V^{xc}(r) ] \psi_{ln}(r) = \epsilon_{ln} \psi_{ln}(r)
\label{psiKB}
\end{eqnarray}
   $V^H$ and $V^{xc}$ are the Hartree and XC potentials for
the pseudo-valence charge density, 
and  we are using atomic units ($e = \hbar = m_e = 1$)
throughout all of this paper.

\begin{figure}[htbp]
\begin{center}
\includegraphics[width=\columnwidth] {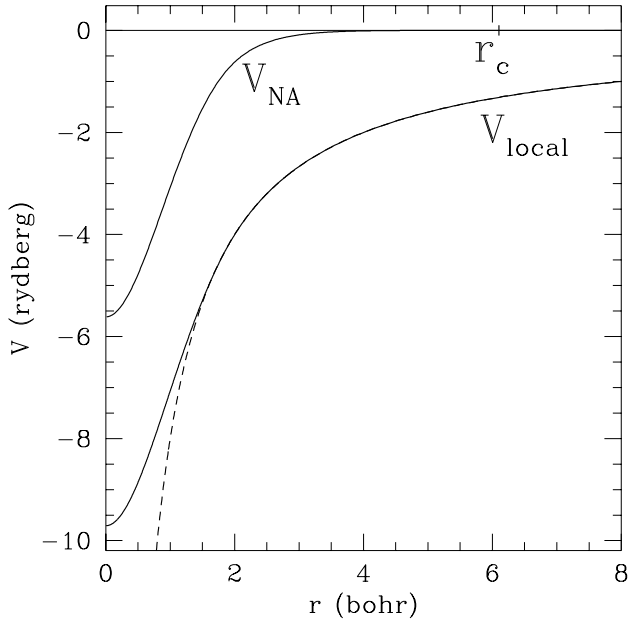}
  \caption{Local pseudopotential for silicon.
        $V_{local}$ is the unscreened local part of the pseudopotential,
     generated as the electrostatic potential produced by a
     localized distribution of positive charge, Eq.~(\ref{rho_local}),
     whose integral is equal to the valence ion charge ($Z=4$ for Si).
        The dashed line is $-Z/r$.
        $V_{NA}$ is the local pseudopotential screened by an electron
     charge distribution, generated by filling the first-$\zeta$
     basis orbitals with the free-atom valence occupations.
        Since these basis orbitals are strictly confined to a radius
     $r^c_{max}$, $V_{NA}$ is also strictly zero beyond that radius.}
  \label{PPlocal}
\end{center}
\end{figure}

   The local part of the pseudopotential $V_{local}(r)$ is in principle
arbitrary, but it must join the semilocal potentials $V_l(r)$ which, 
by construction, all become equal to the (unscreened) all-electron 
potential beyond the pseudopotential core radius  $r_{core}$.
   Thus, $\delta V_l(r) = 0$ for $r > r_{core}$.
   Ramer and Rappe have proposed that $V_{local}(r)$ be optimized for
transferability
 \cite{Ramer-Rappe1999},
but most plane wave schemes make it equal to one of the $V_l(r)$'s
for reasons of efficiency.
   Our case is different because $V_{local}(r)$ is the only
pseudopotential part that needs to be represented in the real space grid,
while the matrix elements of the non-local part $\hat{V}_{KB}$ are cheaply
and accurately calculated by two-center integrals.
   Therefore, we optimize $V_{local}(r)$ for
smoothness, making
it equal to the potential created by a positive charge distribution
of the form~\cite{Vanderbilt1985}
\begin{equation}
\label{rho_local}
\rho^{local}(r) \propto \exp[ -(\sinh(a b r)/\sinh(b))^2],
\end{equation}
where $a$ and $b$ are chosen to provide simultaneously optimal
real-space localization and reciprocal-space convergence\cite{Vlocal1}.
After some numerical tests we have taken $b=$1
and $a=$1.82/$r_{core}$.
   Figure~\ref{PPlocal} shows $V_{local}(r)$ for silicon.

   Since $V_l(r)=V_{local}(r)$ outside $r_{core}$, $\chi^{KB}_{ln}(r)$
is strictly zero beyond that radius, irrespective of the value of
$\epsilon_{ln}$~\cite{Vlocal2}.
Generally it is sufficient to have a single projector $\chi^{KB}_{lm}$
for each angular momentum (i.e. a single term in the sum on $n$).
   In this case we follow the normal practice of making $\epsilon_{ln}$
equal to the valence atomic eigenvalue $\epsilon_l$,
and the function $\varphi_{l}(r)$ in Eq.~\ref{xiKB} is
identical to the corresponding eigenstate $\psi_{l}(r)$.
   In some cases, particularly for alkaline metals, alkaline earths,
and transition metals of the first few columns, we have sometimes
found it necessary to include the semicore states together with the 
valence states \cite{ionicPS}.
   In these cases, we also include two independent KB projectors, 
one for the semicore and one for the valence states.
   However, our pseudopotentials are still norm-conserving rather than
``ultrasoft''
 \cite{Vanderbilt1990}.
   This is because, in our case, it is only the electron density
 that needs to be accurately represented in a real-space grid, 
rather than each wavefunction.
   Therefore, the ultrasoft pseudopotential formalism does not
imply in {\sc Siesta} the same savings as it does in PW schemes.
   Also, since the non-local part of the pseudopotential is a relatively
cheap operator within {\sc Siesta}, we generally (but not necessarily) 
use a larger than usual value of $l^{KB}_{max}$ in Eq.~(\ref{VKB}), 
making it one unit larger than the $l_{max}$ of the basis functions.

\section{Basis set}

   Order-$N$ methods rely heavily on the sparsity of the Hamiltonian and 
overlap matrices.
   This sparsity requires either the neglect of matrix elements that are small enough
or the use of strictly confined basis orbitals, {\it i.e.}, orbitals that are 
zero beyond a certain radius
\cite{Sankey-Niklewski1989}.
   We have adopted this latter approach because it keeps the energy
strictly variational, thus facilitating the test of the convergence
with respect to the radius of confinement. 
   Within this radius, our atomic basis orbitals are products of a 
numerical radial function times a spherical harmonic.
   For atom $I$, located at ${\bf R}_I$,
\begin{equation}
\label{phi}
   \phi_{Ilmn}({\bf r}) = \phi_{Iln}(r_I) Y_{lm}(\hat{\bf r}_I)
\end{equation}
where ${\bf r}_I = {\bf r} - {\bf R}_I$, $r = |{\bf r}|$ and 
$\hat{\bf r} = {\bf r}/r$.
   The angular momentum (labelled by $l,m$) may be arbitrarily large and, 
in general, there will be several orbitals (labelled by index $n$)  
with the same angular dependence, but different radial dependence, 
which is conventionally called a `multiple-$\zeta$' basis.
   The radial functions are defined by a cubic spline interpolation
 \cite{NumericalRecipes}
from the values given on a fine radial mesh.
   Each radial function may have a different cutoff radius and,
up to that radius, its shape is completely free and can be 
introduced by the user in an input file.
   In practice, it is also convenient to have an automatic 
procedure to generate sufficiently good basis sets.
   We have developed several such automatic procedures, and we will
describe here one of them for completeness, even though we stress
that the generation of the basis set, like that of the pseudopotential
is to a large extent up to the user and independent of the {\sc Siesta}
method itself.

   In the case of a minimal (single-$\zeta$) basis set, we have found 
convenient and efficient the method of Sankey and Niklewski 
\cite{Sankey-Niklewski1989,SanchezPortal1996}.
   Their basis orbitals are the eigenfunctions of the (pseudo)atom
within a spherical box (although the radius of the box may be different
for each orbital, see below).
   In other words, they are the (angular-momentum-dependent)
numerical eigenfunctions $\phi_l(r)$ of the atomic pseudopotential
$V_l(r)$, for an energy $\epsilon_l + \delta \epsilon_l$ chosen so 
that the first node occurs at the desired cutoff radius $r^c_l$:
\begin{equation}
\label{SankeyBasis}
  \left( -\frac{1}{2r} \frac{d^2}{d r^2} r
    + \frac{l(l+1)}{2r^2} + V_l(r) \right) \phi_l(r) = 
    (\epsilon_l + \delta \epsilon_l) \phi_l(r)
\end{equation}
with $\phi_l(r^c_l) = 0$ (we omit indices $I$ and $n$ here for simplicity).
   In order to obtain a well balanced basis, in which the effect of the 
confinement is similar for all the orbitals, it is usually better
to fix a common `energy shift' $\delta \epsilon$, rather than a common 
radius $r^c$, for all the atoms and angular momenta.
   This means that the orbital radii depend on the atomic species and 
angular momentum.

   One obvious possibility for multiple-$\zeta$ bases is to use  
pseudopotential eigenfunctions with an increasing number of nodes
\cite{SanchezPortal1996}.
   They have the virtue of being orthogonal and asymptotically complete. 
   However, the efficiency of this kind of basis depends on the radii
of confinement of the different orbitals, since the excited states
of the pseudopotential are usually unbound. 
   Thus, in practice we have found this procedure rather inefficient.
   Another possibility is to use the atomic eigenstates for different
ionization states
 \cite{Lippert-Hutter-Parrinello1996}.
   We have implemented a different scheme
 \cite{Artacho1999}, 
based on the `split-valence' method which is standard in quantum chemistry
 \cite{Huzinaga1984}.
   In that method, the first-$\zeta$ basis orbitals are `contracted' (i.e. fixed) 
linear combinations of Gaussians, determined either variationally or
by fitting numerical atomic eigenfunctions.
   The second-$\zeta$ orbital is then one of the Gaussians (generally the
slowest-decaying one) which is `released' or `split' from the contracted
combination.
   Higher-$\zeta$ orbitals are generated in a similar way by releasing
more Gaussians.
   Our scheme adapts this split-valence method to our numerical orbitals.
   Following the same spirit, our second-$\zeta$ functions 
$\phi_l^{2\zeta}(r)$ have the same tail as the first-$\zeta$ orbitals 
$\phi_l^{1\zeta}(r)$ but change to a simple polynomial behaviour inside 
a `split radius' $r^s_l$:
\begin{equation}
\label{split-valence}
   \phi_l^{2\zeta}(r) = 
      \left\{ 
         \begin{array}{ll}
            r^l (a_l - b_l r^2) & \mbox{if $r < r^s_l$}    \\
            \phi_l^{1\zeta}(r)  & \mbox{if $r \geq r^s_l$}
         \end{array}
      \right. 
\end{equation}
where $a_l$ and $b_l$ are determined by imposing the continuity of value 
and slope at $r^s_l$.
   These orbitals therefore combine the decay of the atomic eigenfunctions
with a smooth and featureless behaviour inside $r^s_l$.
   We have found it convenient to set the radius $r^s_l$ by fixing the norm
of $\phi_l^{1\zeta}$ in $r>r^s_l$.
   We have found empirically that a reasonable value for this 
 `split-norm' is $\sim 0.15$.
   Actually, instead of $\phi_l^{2\zeta}$ thus defined, we use
$\phi_l^{1\zeta} - \phi_l^{2\zeta}$, which is zero beyond  $r^s_l$,
to reduce the number of nonzero matrix elements, without any loss 
of variational freedom.

   To achieve well converged results, in addition to the atomic valence 
orbitals, it is generally necessary to also include polarization orbitals,
to account for the deformation induced by bond formation.
   Again, using pseudoatomic orbitals of higher angular momentum is
frequently unsatisfactory, because they tend to be too extended, or 
even unbound.
   Instead, consider a valence pseudoatomic orbital
$\phi_{lm}({\bf r}) = \phi_l(r) Y_{lm}(\hat{\bf r})$, 
such that there are no valence orbitals with angular momentum $l+1$.
   To polarize it,
we apply a small electric field ${\mathcal E}$ in the $z$-direction.
   Using first-order perturbation theory
\begin{equation}
  (H-E) \delta \phi = -(\delta H -\delta E) \phi,
\label{(H-E)delPsi}
\end{equation}
where 
$\delta H ={\mathcal E} z$ 
and 
$\delta E = \langle \phi | \delta H | \phi \rangle = 0$
because $\delta H$ is odd.
   Selection rules imply that the resulting perturbed orbital will only
have components with $l' = l \pm 1, m'=m$:
\begin{eqnarray}
   \delta H \phi_{lm}({\bf r}) 
   & = & \left( {\mathcal E} r \cos(\theta) \right)
            \left( \phi_l(r) Y_{lm}(\hat{\bf r}) \right)
           \nonumber \\
   & = & {\mathcal E} r \phi_l(r) ( c_{l-1} Y_{l-1,m} + c_{l+1} Y_{l+1,m} )
\label{delHpsi}
\end{eqnarray}
and
\begin{equation}
   \delta \phi_{lm}({\bf r}) = \varphi_{l-1}(r) Y_{l-1,m}(\hat{\bf r}) +
                               \varphi_{l+1}(r) Y_{l+1,m}(\hat{\bf r}).
\label{delPsi}
\end{equation}
   Since in general there will already be orbitals with angular 
momentum $l-1$ in the basis set, we select the $l+1$ component
by substituting (\ref{delHpsi}) and (\ref{delPsi}) in (\ref{(H-E)delPsi}),
multiplying by $Y^*_{l+1,m}(\hat{\bf r})$ and integrating over 
angular variables.
   Thus we obtain the equation
\begin{eqnarray}
   [ -\frac{1}{2r} \frac{d^2}{d r^2} r
        & + & \frac{(l+1)(l+2)}{2r^2} \nonumber \\
        & + & V_l(r) - E_l ]
   \varphi_{l+1}(r) = - r \phi_l(r)
\label{polEq}
\end{eqnarray}
where we have also eliminated the factors ${\mathcal E}$ and $c_{l+1}$, 
which affect only the normalization of $\varphi_{l+1}$.
   The polarization orbitals are then added to the basis set:
$\phi_{l+1,m}({\bf r}) = N \varphi_{l+1}(r) Y_{l+1,m}(\hat{\bf r})$,
where $N$ is a normalization constant.

\begin{figure}
\includegraphics[width=\columnwidth] {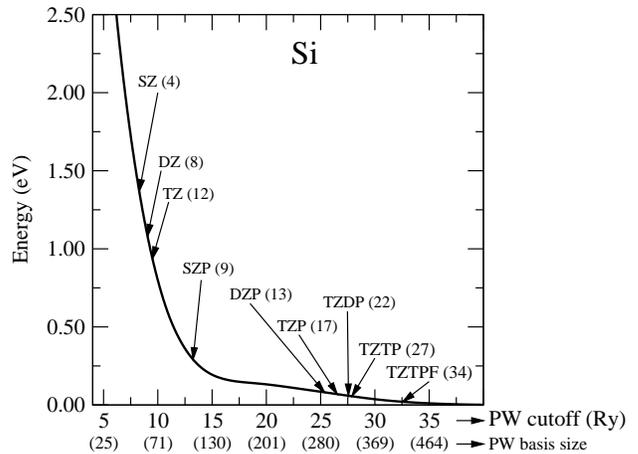}
  \caption{\label{PWcomparison}
                  Comparison of convergence of the total energy with respect 
               to the sizes of a plane wave basis set and of the LCAO
               basis set used by {\sc Siesta}.
                  The curve shows the total energy per atom of silicon versus
               the cutoff of a plane wave basis, calculated with a program
               independent of {\sc Siesta}, which uses the same pseudopotential.
                  The arrows indicate the energies obtained with different LCAO
               basis sets, calculated with {\sc Siesta}, and the plane wave cutoffs
               that yield the same energies.
                  The numbers in parentheses indicate the basis sizes, i.e. the
               number of atomic orbitals or plane waves of each basis set.
               SZ: single-$\zeta$ (valence $s$ and $p$ orbitals);
               DZ: double-$\zeta$;
               TZ: triple-$\zeta$;
               DZP: double-$\zeta$ valence orbitals plus single-$\zeta$
                       polarization $d$ orbitals;
               TZP: triple-$\zeta$ valence plus single-$\zeta$  polarization;
               TZDP: triple-$\zeta$ valence plus double-$\zeta$  polarization;
               TZTP: triple-$\zeta$ valence plus triple-$\zeta$  polarization;
               TZTPF: same as TZTP plus extra single-$\zeta$ polarization
                          $f$ orbitals.
               }
\end{figure}

   We have found that the previously described procedures generate reasonable
minimal single-$\zeta$ (SZ) basis sets, appropriate for semiquantitative simulations, 
and double-$\zeta$ plus polarization (DZP) basis sets that yield high quality 
results for most of the systems studied.
   We thus refer to DZP as the `standard' basis, because it usually
represents a good balance between well converged results and a reasonable
computational cost.
   In some cases (typically alkali and some transition metals), semicore states 
also need to be included for good quality results.
   More recently
 \cite{Junquera2001:bases}, 
we have obtained extremely efficient basis sets optimized variationally
in molecules or solids.
   Figure \ref{PWcomparison} shows the performance of these atomic basis 
sets compared to plane waves, using the same pseudopotentials and 
geometries.
   It may be seen that the SZ bases are comparable to planewave cutoffs
typically used in Car-Parrinello molecular dynamics simulations, while
DZP sets are comparable to the cutoffs used in geometry relaxations
and energy comparisons.
   As expected, the LCAO is far more efficient, tipically by a factor
of 10 to 20, in terms of number of basis orbitals.
   This efficiency must be balanced against the faster
algorithms available for plane waves, and our main motivation for
using an LCAO basis is its suitability for ${\cal O}(N)$ methods.
   Still, we have generally found that, even without using the 
${\cal O}(N)$ functional, {\sc Siesta} is considerably faster than a plane 
wave calculation of similar quality.

\begin{figure}[htbp]
\includegraphics[width=\columnwidth] {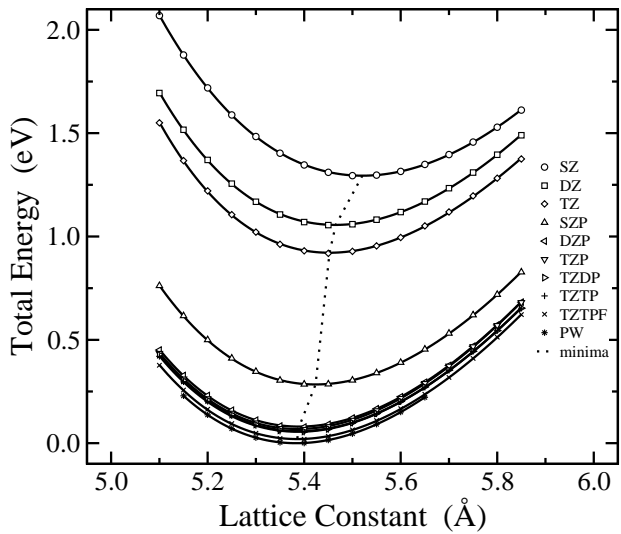}
   \caption{   Total energy per atom versus lattice constant for bulk silicon,
                using different basis sets, noted as in Fig.~\ref{PWcomparison}.
                   PW refers to a very well converged (50 Ry cutoff) plane wave
                calculation.
                   The dotted line joins the minima of the different curves.
               }
\label{BasisConvFig}
\end{figure}

   Figure \ref{BasisConvFig} shows the convergence of the 
total energy curve of silicon, as a function of lattice parameter,
for different basis sizes, and table \ref{BasisConvTable} summarizes
the same information numerically.
   It can be seen that the `standard' DZP basis offers already quite
well converged results, comparable to those used in practice in
most plane wave calculations.

\begin{table} 
   \caption[ ]{  Comparisons of the lattice constant $a$, bulk modulus $B$,
                  and cohesive energy $E_c$ for bulk Si, obtained with
                  different basis sets.
                      The basis notation is as in Fig.~\ref{PWcomparison}.
                      PW refers to a 50 Ry-cutoff plane wave calculation.
                      The LAPW results were taken from 
                  ref.~\onlinecite{Filippi1994},  and the experimental values from 
                  ref.~\onlinecite{Kittel1986}.
                  }
\begin{center}
\begin{tabular}{lccc}
Basis & $a$ (\AA) & $B$ (GPa) & $E_c$ (eV) \\
\hline
SZ      & 5.521 & 88.7 & 4.722   \\
DZ      & 5.465 & 96.0 & 4.841   \\
TZ      & 5.453 & 98.4 & 4.908   \\
SZP     & 5.424 & 97.8 & 5.227   \\
DZP     & 5.389 & 96.6 & 5.329   \\
TZP     & 5.387 & 97.5 & 5.335   \\
TZDP    & 5.389 & 96.0 & 5.340   \\
TZTP    & 5.387 & 96.0 & 5.342   \\
TZTPF   & 5.385 & 95.4 & 5.359   \\
\hline
PW      & 5.384 & 95.9 & 5.369   \\
LAPW    & 5.41  & 96   & 5.28    \\
Expt.   & 5.43  & 98.8 & 4.63    \\
\hline
\end{tabular}
\end{center}
\label{BasisConvTable}
\end{table}

\begin{figure}[htbp]
\includegraphics[width=\columnwidth] {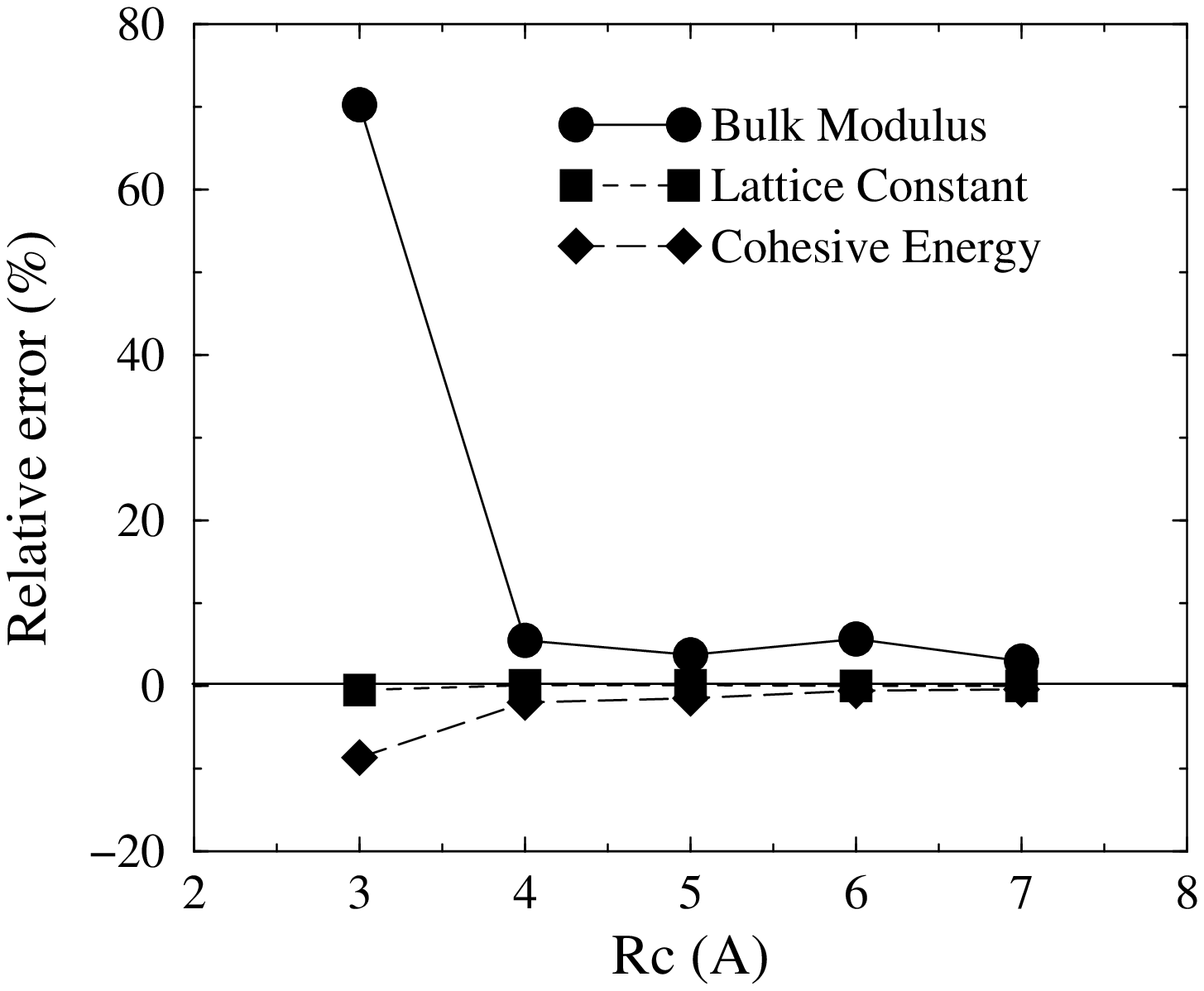}
\caption{   Dependence of the lattice constant, bulk modulus, and 
            cohesive energy of bulk silicon with the cutoff radius of the 
            basis orbitals.
                The $s$ and $p$ orbital radii have been made equal in  
            this case, to simplify the plot.
                PW refers to a well converged plane wave calculation
            with the same pseudopotential.
            }
\label{rcConv}
\end{figure}

   Figure \ref{rcConv} shows the dependence of the lattice constant,
bulk modulus, and cohesive energy of bulk silicon with the range of
the basis orbitals.
   It shows that a cutoff radius of 3 \AA~ for both $s$ and $p$ orbitals
yields already very well converged results, specially when using a
`standard' DZP basis.

\section{Electron Hamiltonian}

   Within the non-local-pseudopotential approximation, 
the standard Kohn-Sham one-electron Hamiltonian may be written as
\begin{equation}
\label{Hinitial}
   \hat{H} = \hat{T} + \sum_I V_I^{local}({\bf r}) + 
      \sum_I \hat{V}_I^{KB} + V^{H}({\bf r}) + V^{xc}({\bf r}) 
\end{equation}
where $\hat{T}=-\frac{1}{2} \nabla^2$ is the kinetic energy operator,
$I$ is an atom index, $V^{H}({\bf r})$ and $V^{xc}({\bf r})$ 
are the total Hartree and XC  potentials, and
$V_I^{local}({\bf r})$ and $\hat{V}_I^{KB}$
are the local and non-local (Kleinman-Bylander) parts of the 
pseudopotential of atom $I$.

   In order to eliminate the long range of $V_I^{local}$, we screen
it with the potential $V_I^{atom}$, created by an atomic electron density 
$\rho_I^{atom}$, constructed by populating the basis 
functions with appropriate valence atomic charges. 
   Notice that, since the atomic basis orbitals are zero beyond 
the cutoff radius $r_I^c = \max_l(r^c_{Il})$, 
the screened `neutral-atom' (NA) potential 
$V_I^{NA} \equiv V_I^{local} + V_I^{atom}$ 
is also zero beyond this radius
 \cite{Sankey-Niklewski1989}
(see Fig.~\ref{PPlocal}).
   Now let $\delta \rho({\bf r})$ be the difference between
the self-consistent electron density $\rho({\bf r})$ and the sum
of atomic densities 
$\rho^{atom} = \sum_I \rho_I^{atom}$, 
and let $\delta V^H({\bf r})$
be the electrostatic potential generated by $\delta \rho({\bf r})$,
which integrates to zero and is usually much smaller than $\rho({\bf r})$.
   Then the total Hamiltonian may be rewritten as
\begin{equation}
\label{Hfinal}
\hat{H} = \hat{T} + \sum_I \hat{V}_I^{KB} + \sum_I V_I^{NA}({\bf r}) +
    \delta V^{H}({\bf r}) + V^{xc}({\bf r}) 
\end{equation}
   The matrix elements of the first two terms involve only two-center 
integrals which are calculated in reciprocal space and tabulated as a 
function of interatomic distance.
   The remaining terms involve potentials which are calculated on a
three-dimensional real-space grid.
   We consider these two approaches in detail in the following sections.

\section{Two-center integrals}
\label{Two-center}

   The overlap matrix and the largest part of
the Hamiltonian matrix elements are given by two-center integrals
 \cite{three_center}.
   We calculate these integrals in Fourier space, as proposed by
Sankey and Niklewski 
 \cite{Sankey-Niklewski1989},
but we use some implementation details explained in this section.
   Let us consider first overlap integrals of the form
\begin{equation}
\label{two_center}
   S({\bf R}) \equiv \langle \psi_1 | \psi_2 \rangle 
      = \int \psi^*_1({\bf r}) \psi_2({\bf r-R}) d{\bf r},
\end{equation}
where the integral is over all space and
$\psi_1, \psi_2$ may be basis functions $\phi_{lmn}$, 
KB pseudopotential projectors $\chi_{lmn}$, or more complicated 
functions centered on the atoms.
   The function $S({\bf R})$ can be seen as a convolution:
we take the Fourier transform
\begin{equation}
\label{Fourier_transform}
   \psi({\bf k}) = \frac{1}{(2\pi)^{3/2}}
      \int \psi({\bf r}) e^{-i{\bf kr}} d{\bf r} 
\end{equation}
where we use the same symbol $\psi$ for $\psi({\bf r})$ and
$\psi({\bf k})$, as its meaning is clear from the different arguments.
   We also use the planewave expression of Dirac's delta function,
$\int e^{i({\bf k'-k}){\bf r}} d{\bf r} = (2\pi)^3 \delta({\bf k'-k})$,
to find the usual result that the Fourier transform of a convolution
in real space is a simple product in reciprocal space:
\begin{equation}
\label{S_transform}
   S({\bf R}) = \int \psi_1^*({\bf k}) \psi_2({\bf k})
                     e^{-i{\bf kR}} d{\bf k} 
\end{equation}
   Let us assume now that the functions $\psi({\bf r})$ can
be expanded exactly with a finite number of spherical harmonics:
\begin{equation}
\label{psi_factorization}
   \psi({\bf r}) = \sum_{l=0}^{l_{max}} \sum_{m=-l}^l
                   \psi_{lm}(r) Y_{l m}(\hat{\bf r}),
\end{equation}
\begin{equation}
\label{psi_lm}
   \psi_{lm}(r) = \int_0^{\pi} \sin\theta d\theta \int_0^{2\pi} d\varphi
      Y^*_{l m}(\theta,\varphi) \psi(r,\theta,\varphi).
\end{equation}
   This is clearly true for basis functions and KB projectors, which
contain a single spherical harmonic, and also for functions like
$x\psi({\bf r})$, which appear in dipole matrix elements.
   We now substitute in (\ref{Fourier_transform}) the expansion of a 
plane wave in spherical harmonics
 \cite{Jackson1962}
\begin{equation}
\label{PWexp}
   e^{i{\bf kr}} = \sum_{l=0}^{\infty} \sum_{m=-l}^l
      4 \pi i^l j_l(kr) Y^*_{lm}(\hat{\bf k}) Y_{lm}(\hat{\bf r}),
\end{equation}
to obtain
\begin{equation}
\label{psik}
   \psi({\bf k}) = \sum_{l=0}^{l_{max}} \sum_{m=-l}^l
                   \psi_{lm}(k) Y_{l m}(\hat{\bf k}),
\end{equation}
\begin{equation}
\label{psikr}
   \psi_{lm}(k) = \sqrt{\frac{2}{\pi}} (-i)^{l}
      \int_0^{\infty} r^2 dr j_l(kr) \psi_{lm}(r).
\end{equation}
   Substituting now (\ref{psik}) and (\ref{PWexp}) into 
(\ref{S_transform}) we obtain
\begin{equation}
\label{S_final}
   S({\bf R}) = \sum_{l=0}^{2l_{max}} \sum_{m=-l}^l
       S_{lm}(R) Y_{lm}(\hat{\bf R})
\end{equation}
where
\begin{equation}
\label{Slm}
   S_{lm}(R) = \sum_{l_1 m_1} \sum_{l_2 m_2}
       G_{l_1 m_1, l_2 m_2, l m} S_{l_1 m_1, l_2 m_2, l}(R),
\end{equation}
\begin{eqnarray}
\label{Gaunt}
   G_{l_1 m_1, l_2 m_2, l m} 
     && = \int_0^{\pi} \sin\theta d\theta \int_0^{2\pi} d\varphi
           \nonumber \\
     && Y^*_{l_1 m_1}(\theta,\varphi) Y_{l_2 m_2}(\theta,\varphi)
              Y^*_{l m}(\theta,\varphi),
\end{eqnarray}
\begin{eqnarray}
\label{Sl1l2l}
   S_{l_1 m_1, l_2 m_2, l}(R) 
     & = & 4 \pi i^{l_1-l_2-l} \int_0^\infty k^2 dk
           j_l(kR) \nonumber \\
     &\times& i^{-l_1} \psi^*_{1, l_1 m_1}(k) 
              i^{l_2} \psi_{2, l_2 m_2}(k),
\end{eqnarray}
   Notice that $i^{-l_1} \psi_1(k), i^{l_2} \psi_2(k)$, and
$i^{l_1-l_2-l}$ are all real, since $l_1-l_2-l$ is even for all
$l$'s for which $G_{l_1 m_1, l_2 m_2, l m} \neq 0$.
   The Gaunt coefficients $G_{l_1 m_1, l_2 m_2, l m}$ can be obtained 
by recursion from Clebsch-Gordan coefficients
 \cite{Sankey-Niklewski1989}.
   However, we use real spherical harmonics 
for computational efficiency:
\begin{equation}
\label{Ylm}
   Y_{lm}(\theta,\varphi) = C_{lm} P_l^m(\cos\theta) \times
      \left\{ 
         \begin{array}{ll}
            \sin(m\varphi) & \mbox{if $m < 0$}    \\
            \cos(m\varphi) & \mbox{if $m \geq 0$}
         \end{array}
      \right. 
\end{equation}
where $P_l^m(z)$ are the associated Legendre polynomials and $C_{lm}$ 
normalization constants
 \cite{NumericalRecipes}.
   This does not affect the validity of any of previous equations,
but it modifies the value of the Gaunt coefficients.
   Therefore, we find it is simpler and more general to calculate
$G_{l_1 m_1, l_2 m_2, l m}$ directly from Eq.~(\ref{Gaunt}).
   To do this, we use a Gaussian quadrature
 \cite{NumericalRecipes}
\begin{equation}
\label{cuadrature}
   \int_0^{\pi} \sin\theta d\theta \int_0^{2\pi} d\varphi \rightarrow
      4 \pi \frac{1}{N_{\theta}} \sum_{i=1}^{N_{\theta}} w_i \sin\theta_i
            \frac{1}{N_{\varphi}} \sum_{j=1}^{N_{\varphi}}
\end{equation}
with $N_{\varphi}=1+3l_{max}, N_{\theta}=1+\mbox{int}(3l_{max}/2)$, and
the points $\cos\theta_i$ and weights $w_i$ are calculated as described
in ref.~\onlinecite{NumericalRecipes}.
   This quadrature is exact in equation (\ref{Gaunt}) for spherical
harmonics $Y_{lm}$ (real or complex) of $l \leq l_{max}$, and it can be
used also to find the expansion of $\psi({\bf r})$ in spherical
harmonics (eq.~(\ref{psi_lm})).

   The coefficients $G_{l_1 m_1, l_2 m_2, l m}$ are universal and
they can be calculated and stored once and for all.
   The functions $S_{l_1 m_1, l_2 m_2, l}(R)$ depend, of course, on the
functions $\psi_{1,2}({\bf r})$ being integrated. 
   For each pair of
functions, they can be calculated and stored in a fine radial grid $R_i$,
up to the maximum distance $R_{max}=r^c_1+r^c_2$ at which 
$\psi_1$ and $\psi_2$ overlap.
   Their value at an arbitrary distance $R$ can then be obtained very
accurately using a spline interpolation.

   Kinetic matrix elements 
$T({\bf R}) \equiv \langle \psi^*_1 | -\frac{1}{2}\nabla^2 | \psi_2 \rangle$
can be obtained in exactly the same way, except for an extra factor $k^2$
in Eq.~(\ref{Sl1l2l}):
\begin{eqnarray}
\label{Tl1l2l}
   T_{l_1 m_, l_2 m_2, l}(R) 
     & = & 4 \pi i^{l_1-l_2-l} \int_0^\infty \frac{1}{2} k^4 dk
           j_l(kR) \nonumber \\
     &\times& i^{-l_1} \psi^*_{1, l_1 m_1}(k) 
              i^{l_2} \psi_{2, l_2 m_2}(k).
\end{eqnarray}
   Since we frequently use basis orbitals with a kink
 \cite{Sankey-Niklewski1989},
we need rather fine radial grids to obtain accurate kinetic matrix
elements, and we typically use grid cutoffs of more than 2000 Ry for 
this purpose.
   Once obtained, the fine grid does not penalize the execution time, 
because the interpolation effort is independent of the number of
grid points.
   It also affects very marginally the storage requirements,
because of the one-dimensional character of the tables.
   However, even though it needs to be done only once, the calculation 
of the radial integrals (\ref{psikr}), (\ref{Sl1l2l}), and (\ref{Tl1l2l}) is not 
negligible if performed unwisely.
   We have developed a special fast radial Fourier transform for
this purpose, as explained in appendix \ref{RadFFT}.

   Dipole matrix elements, such as $\langle \psi_1 | x | \psi_2 \rangle$,
can also be obtained easily by defining a new function
$\chi_1({\bf r}) \equiv x \psi_1({\bf r})$, expanding it using
(\ref{psi_lm}), and computing $\langle \chi_1 | \psi_2 \rangle$
as explained above (with the precaution of using $l_{max}+1$ instead of 
$l_{max}$).

\section{Grid integrals}

   The matrix elements of the last three terms  of Eq.~(\ref{Hfinal}) 
involve potentials which are calculated on a real-space grid.
   The fineness of this grid is controlled by a `grid cutoff' $E_{cut}$:
the maximum kinetic energy of the planewaves that can be represented
in the grid without aliasing \cite{densityEcut}.
   The short-range screened pseudopotentials  $V^{NA}_I({\bf r})$ 
in (\ref{Hfinal}) are tabulated as a function of the distance to atoms $I$
and easily interpolated at any desired grid point.
   The last two terms require the calculation of the electron density
on the grid.
   Let $\psi_i({\bf r})$ be the Hamiltonian eigenstates, expanded in
the atomic basis set 
\begin{equation}
\label{psi}
   \psi_i({\bf r}) = \sum_\mu \phi_\mu({\bf r}) c_{\mu i} , 
\end{equation}
where $c_{\mu i} = \langle \tilde{\phi}_\mu | \psi_i \rangle$ and
$\tilde{\phi}_\mu$ is the dual orbital of $\phi_\mu$:
$\langle \tilde{\phi}_\mu | \phi_\nu \rangle = \delta_{\mu \nu}$.
   We use the compact index notation $\mu \equiv \{Ilmn\}$ for 
the basis orbitals, Eq.~(\ref{phi}).
   The electron density is then
\begin{equation}
\label{rho}
   \rho({\bf r}) = \sum_i n_i | \psi_i({\bf r}) |^2 
\end{equation}
where $n_i$ is the occupation of state $\psi_i$.
   If we substitute (\ref{psi}) into (\ref{rho}) and define a density matrix
\begin{equation}
\label{DM}
   \rho_{\mu \nu} = \sum_i  c_{\mu i} n_i c_{i \nu},
\end{equation}
where $c_{i \nu} \equiv c^*_{\nu i}$, the electron density can 
be rewritten as
\begin{equation}
\label{DM2rho}
   \rho({\bf r}) = 
      \sum_{\mu \nu} \rho_{\mu \nu} \phi^*_\nu({\bf r}) \phi_\mu({\bf r})
\end{equation}
   We use the notation $\phi^*_\mu$ for generality, despite our use
of real basis orbitals in practice.
   Then, to calculate the density at a given grid point, we first find 
all the atomic basis orbitals, Eq.~(\ref{phi}), at that point, 
interpolating the radial part from numerical tables, and then we use 
(\ref{DM2rho}) to calculate the density.
   Notice that only a small number of basis orbitals are
non-zero at a given grid point, so that the calculation of the density 
can be performed in ${\cal O}(N)$ operations, once $\rho_{\mu \nu}$ is known.
The storage of the orbital values at the grid points can be one of the
most expensive parts of the program in terms of memory usage. Hence, an
option is included to calculate and use these terms on the fly, in the
the spirit of a direct-SCF calculation.
   The calculation of $\rho_{\mu \nu}$ itself with Eq.~(\ref{DM}) does
not scale linearly with the system size, requiring instead the use
of special ${\cal O}(N)$ techniques to be described below.
   However, notice that in order to calculate the density, only the matrix elements
$\rho_{\mu \nu}$ for which $\phi_\mu$ and $\phi_\nu$ overlap are required,
and they can therefore be stored as a sparse matrix of ${\cal O}(N)$ size.
   Once the valence density is available in the grid, we add to it, if
necessary, the non-local core correction 
 \cite{Louie-Froyen-Cohen1982:NLCC},
a spherical charge density intended to simulate the atomic cores, which 
is also interpolated from a radial grid.
   With it, we find the exchange
and correlation potential $V^{xc}({\bf r})$, trivially in the LDA and 
using the method described in 
ref.~\onlinecite{Balbas2001}
for the GGA.
   To calculate $\delta V^{H}({\bf r})$, we first find 
$\rho^{atom}({\bf r})$ at the grid points, as a sum of spherical 
atomic densities (also interpolated from a radial grid) and subtract 
it from $\rho({\bf r})$ to find $\delta \rho({\bf r})$.
   We then solve Poisson's equation to obtain $\delta V^{H}({\bf r})$
and find the total grid potential 
$V({\bf r}) = V^{NA}({\bf r}) + \delta V^{H}({\bf r}) + V^{xc}({\bf r})$.
   Finally, at every grid point, we calculate
$V({\bf r}) \phi^*_\mu({\bf r}) \phi_\nu({\bf r}) \Delta {\bf r}^3$
for all pairs $\phi_\mu, \phi_\nu$ which are not zero at that point 
($\Delta {\bf r}^3$ is the volume per grid point) and add
it to the Hamiltonian matrix element $H_{\mu \nu}$.

   To solve Poisson's equation and find  $\delta V^{H}({\bf r})$
we normally  use  fast Fourier transforms in a unit cell
that is either naturally periodic or made artificially
periodic by a supercell construction.
   For neutral isolated molecules, our use of strictly confined basis
orbitals makes it trivial to avoid any direct overlap between
the repeated molecules, and the electric multipole interactions
decrease rapidly with cell size.
   For charged molecules we supress the 
${\bf G}=0$ Fourier component (an infinite constant) of the 
potential created by the excess of charge.
   This amounts to compensating this excess with a uniform 
charge background.
   We then use the method of Makov and Payne
\cite{Makov-Payne1995}
to correct the total energy for the interaction between the
repeated cells.
   Alternatively, we can solve Poisson's equation 
by the multigrid method, using finite differences and fixed
boundary conditions, obtained from the multipole expansion 
of the molecular charge density. 
   This can be done in strictly ${\cal O}(N)$ operations, 
unlike the FFT's, which scale as $N \log N$.
   However, the cost of this operation is
typically negligible and therefore has no influence on the
overall scaling properties of the calculation.

\begin{figure}[htbp]
\includegraphics[width=\columnwidth] {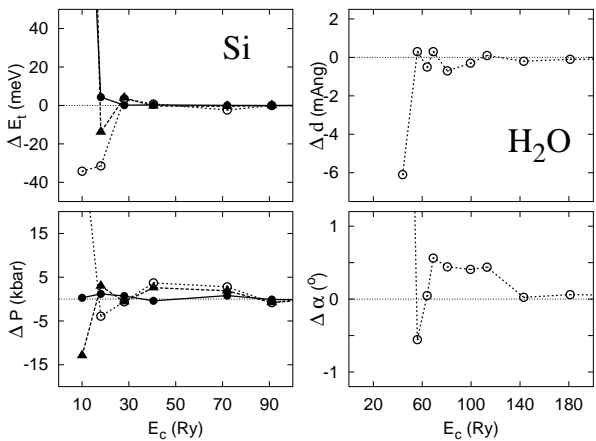}
\caption{   
    (a) Convergence of the total energy and pressure in bulk
           silicon as a function of the energy cutoff $E_{cut}$ 
           of the real space integration mesh.
         Circles and continuous line: using a grid-cell-sampling
           of eight refinement points per original grid point.
         The refinement points are used only in the final calculation,
           not during the self-consistency iteration (see text).
         Triangles: two refinement points per original grid point.
         White circles: no grid-cell-sampling.
   (b) Bond length and angle of the water molecule as a function
          of $E_{cut}$
            }
\label{GridConvSi+H2O}
\end{figure}

\begin{figure}[htbp]
\includegraphics[width=\columnwidth] {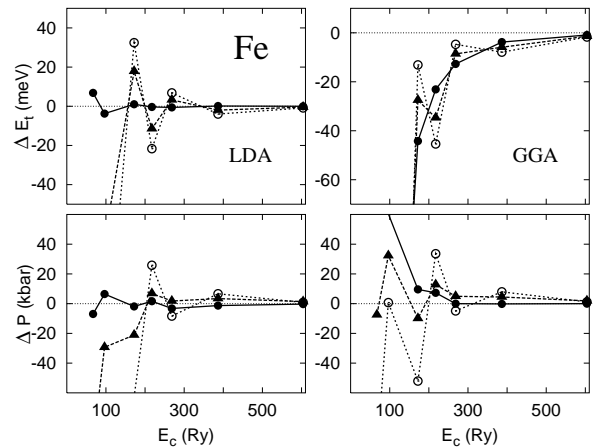}
\caption{   Same as Fig.~\ref{GridConvSi+H2O} for the total
         energy and pressure of bulk iron.
             This is presented as a specially difficult case because
         of the very hard partial core correction ($r_m$ = 0.7 au) 
         required for a correct description of exchange and correlation.
            }
\label{GridConvFe}
\end{figure}

   Figures \ref{GridConvSi+H2O} and \ref{GridConvFe} show the
convergence of different magnitudes with respect to the
energy cutoff of the  integration grid.
   For orthogonal unit cell vectors this is simply, in atomic units,
$E_{cut} = (\pi / \Delta x)^2 / 2$ with $\Delta x$ the grid interval.

\section{Non-collinear spin}

   In the usual case of a normal (collinear) spin polarized system, 
there are two sets of values for
$\psi_i({\bf r}), \rho_{\mu \nu}, \rho({\bf r}), 
V^{xc}({\bf r})$, and $H_{\mu \nu}$, 
one for spin up and another for spin down.
   Thus, the grid calculations can be repeated twice in an almost
independent way: only to calculate $V^{xc}({\bf r})$ need they be 
combined.
   However, in the non-collinear spin case
 \cite{Sandratskii-Guletskii1986,Kubler1988,Oda-Pasquarello-Car1998,
       Postnikov2001}, 
the density at every point is not represented by the up and 
down values, but also by a vector giving the spin direction.
   Equivalently, it may be represented by a local spin density matrix
\begin{equation}
\label{spinDM}
   \rho^{\alpha \beta}({\bf r}) = 
      \sum_i n_i \psi_i^{\beta *}({\bf r}) \psi_i^{\alpha}({\bf r}) =
      \sum_{\mu \nu} \rho_{\mu \nu}^{\alpha \beta} 
         \phi^*_\nu({\bf r}) \phi_\mu({\bf r})
\end{equation}
\begin{equation}
\label{psi-non-collinear}
   \psi_i^\alpha({\bf r}) = \sum_\mu \phi_\mu({\bf r}) c^\alpha_{\mu i}
\end{equation}
\begin{equation}
\label{totalDM}
   \rho^{\alpha \beta}_{\mu \nu} = 
      \sum_i c^{\alpha}_{\mu i} n_i c^{\beta}_{i \nu}
\end{equation}
where $\alpha, \beta$ are spin indices, with up or down values.
   The coefficients $c^{\alpha}_{\mu i}$ are obtained by solving
the generalized eigenvalue problem
\begin{equation}
\label{eigensystem}
   \sum_{\nu \beta} (H^{\alpha \beta}_{\mu \nu} - 
                     E_i S_{\mu \nu} \delta^{\alpha \beta}) 
                     c^{\beta}_{\nu i} = 0
\end{equation}
where $H^{\alpha \beta}_{\mu \nu}$, like $\rho^{\alpha \beta}_{\mu \nu}$,
is a ($2N \times 2N$) matrix, with $N$ the number of basis functions:
\begin{equation}
\label{H-non-collinear}
  H^{\alpha \beta}_{\mu \nu} = \langle \phi_\mu | 
       \hat{T} + \hat{V}^{KB} + V^{NA}({\bf r}) + \delta V^H({\bf r}) + 
       V_{XC}^{\alpha \beta}({\bf r}) | \phi_\nu \rangle.
\end{equation}
   This is in contrast to the collinear spin case, in which the
Hamiltonian and density matrices can be factorized into two $N \times N$
matrices, one for each spin direction.
   To calculate $V_{XC}^{\alpha \beta}({\bf r})$ we first diagonalize
the $2 \times 2$ matrix $\rho^{\alpha \beta}({\bf r})$ at every point, 
in order to find the up and down spin densities 
$\rho^{\uparrow}({\bf r}), \rho^{\downarrow}({\bf r})$
in the direction of the local spin vector.
   We then find $V_{XC}^{\uparrow}({\bf r}), V_{XC}^{\downarrow}({\bf r})$
in that direction, with the usual local spin density functional
 \cite{Perdew-Zunger1981}
and we rotate back $V_{XC}^{\alpha \beta}({\bf r})$ to the original
direction.
   Thus, the grid operations are still basically the same, except that
they need now be repeated three times, for the 
$\uparrow \uparrow, \downarrow \downarrow$ and $\uparrow \downarrow$
components.
   Notice that $\rho^{\alpha \beta}({\bf r})$ and 
$V_{XC}^{\alpha \beta}({\bf r})$ are locally Hermitian, while
$H^{\alpha \beta}_{\mu \nu}$ and 
$\rho^{\alpha \beta}_{\mu \nu}$ are globally Hermitian
($H^{\beta \alpha}_{\nu \mu} = H^{\alpha \beta *}_{\mu \nu}$),
so that their $\downarrow \uparrow$ components can be obtained 
from the $\uparrow \downarrow$ ones.

\section{Brillouin zone sampling}

   Integration of all magnitudes over the Brillouin zone (BZ) is 
essential for small and moderately large unit cells, especially 
of metals.
   Although {\sc Siesta} is designed for large unit cells, in practice it
is very useful, especially for comparisons and checks, to be able to 
also perform calculations efficiently on smaller systems without 
using expensive superlattices.
   On the other hand, an efficient $k$-sampling implementation should
not penalize, because of the required complex arithmetic, the 
$\Gamma$-point calculations used in large cells.
   A solution used in some programs is to have two different 
versions of all or part of the  code, but this poses extra 
maintenance requirements.
   We have dealt with this problem in the following way:
around the unit cell (and comprising itself) we define an auxiliary 
supercell large enough to contain all the atoms whose basis orbitals 
are non-zero at any of the grid points of the unit cell, or which 
overlap with any of the basis orbitals in it.
   We calculate all the non-zero two-center integrals
between the unit cell basis orbitals and the supercell orbitals,
without any complex phase factors.
   We also calculate the grid integrals between {\em all} the supercell 
basis orbitals $\phi_{\mu'}$ and $\phi_{\nu''}$ (primed indices run over 
all the supercell), but {\em within the unit cell only}.
   We accumulate these integrals in the corresponding matrix elements,
thus making use of the relation 
\begin{equation}
   <\phi_{\mu} | V({\bf r}) | \phi_{\nu'} > =
\sum_{(\mu' \nu'') \equiv (\mu \nu')}
 <\phi_{\mu'} | V({\bf r}) f({\bf r}) | \phi_{\nu''}>.
\end{equation}
$f({\bf r})=1$ for ${\bf r}$  within the unit cell and is zero otherwise.
   $\phi_\mu$ is within the unit cell.
   The notation $\mu' \equiv \mu$ indicates that $\phi_{\mu'}$ and
$\phi_\mu$ are equivalent orbitals, related by a lattice vector 
translation.
   $(\mu' \nu'') \equiv (\mu \nu')$ means that the sum
extends over all pairs of supercell orbitals $\phi_{\mu'}$ and 
$\phi_{\nu''}$ such that $\mu' \equiv \mu$, $\nu'' \equiv \nu'$, and
${\bf R}_\mu - {\bf R}_{\nu'} = {\bf R}_{\mu'} - {\bf R}_{\nu''}$.
   Once all the real overlap and Hamiltonian matrix elements are 
calculated, we multiply them,
at every $k$-point by the corresponding phase factors and accumulate 
them by folding the supercell orbital to its unit-cell counterpart.
   Thus
\begin{equation}
\label{Hk}
   H_{\mu \nu}({\bf k}) = \sum_{\nu' \equiv \nu} H_{\mu \nu'} 
      e^{i{\bf k}({\bf R}_{\nu'}-{\bf R}_{\mu})}
\end{equation}
where $\phi_\mu$ and $\phi_\nu$ are within the unit cell.
   The resulting $N \times N$ complex eigenvalue problem,
with $N$ the number of orbitals in the unit cell, is then solved
at every sampled $k$ point, finding the Bloch-state expansion
coefficients $c_{\mu i}({\bf k})$:
\begin{equation}
\label{psi_k}
     \psi_i({\bf k, r}) = \sum_{\mu'} 
      e^{i {\bf kR_{\mu'}}}\phi_{\mu'}({\bf r}) c_{\mu' i}({\bf k})
\end{equation}
where the sum in $\mu'$ extends to all basis orbitals in space,
$i$ labels the different bands, 
$c_{\mu' i} = c_{\mu i}$ if $\mu' \equiv \mu$,
and $\psi_i({\bf k, r})$ is normalized
in the unit cell.

   The electron density is then
\begin{eqnarray}
\label{rho_k}
   \rho({\bf r}) 
      & = & \sum_i \int_{BZ} n_i({\bf k}) |\psi_i({\bf k,r})|^2 d{\bf k} 
            \nonumber \\
      & = &  \sum_{\mu' \nu'} \rho_{\mu' \nu'} 
             \phi^*_{\nu'}({\bf r}) \phi_{\mu'}({\bf r})
\end{eqnarray}
where the sum is again over all basis orbitals in space, and the
density matrix
\begin{equation}
\label{DM_k}
   \rho_{\mu \nu} = \sum_i \int_{BZ} 
                    c_{\mu i}({\bf k}) n_i({\bf k}) c_{i \nu}({\bf k})
                e^{i{\bf k}({\bf R_{\nu}-R_{\mu}})} d{\bf k} 
\end{equation}
is real (for real $\phi_\mu$'s) and periodic, i.e. 
$\rho_{\mu \nu} = \rho_{\mu' \nu'}$ if $(\nu,\mu )\equiv (\nu', \mu')$ 
(with `$\equiv$' meaning again  `equivalent by translation').
   Thus, to calculate the density at a grid point of the unit cell,
we simply find the sum (\ref{rho_k}) over all the pairs of orbitals
$\phi_\mu, \phi_\nu$ in the supercell that are non-zero at that point.

   In practice, the integral in (\ref{DM_k}) is performed in a finite,
uniform grid of the Brillouin zone.
   The fineness of this grid is controlled by a $k$-grid cutoff
$l_{cut}$, a real-space radius which plays a role equivalent to the 
planewave cutoff of the real-space grid
 \cite{Moreno-Soler1992}.
   The origin of the $k$-grid may be displaced from ${\bf k} = 0$
in order to decrease the number of inequivalent $k$-points
 \cite{Monkhorst-Pack1976}.

   If the unit cell is large enough to allow a $\Gamma$-point-only
calculation, the multiplication by phase factors is skipped
and a single real-matrix eigenvalue problem is solved
(in this case, the real matrix elements are accumulated directly in 
the first stage, if multiple overlaps occur).
   In this way, no complex arithmetic penalty occurs, and the 
differences between $\Gamma$-point and $k$-sampling  are limited
to a very small section of the code, while all the two-center and
grid integrals use always the same real-arithmetic code.

\section{Total energy}

   The Kohn-Sham 
 \cite{Kohn-Sham1965}
total energy can be written as a sum of a band-structure (BS) energy
plus some correction terms, sometimes called `double count' corrections.
   The BS term is the sum of the energies of the occupied states 
$\psi_i$:
\begin{equation}
\label{E_BS}
   E^{BS} = \sum_i n_i \langle \psi_i | \hat{H} | \psi_i \rangle
          = \sum_{\mu \nu} H_{\mu \nu} \rho_{\nu \mu} 
          = \mbox{Tr}(H \rho)
\end{equation}
where spin and $k$-sampling notations are omitted here for simplicity.
   At convergence, the $\psi_i$'s are simply the eigenvectors of the 
Hamiltonian, but it is important to realize that the Kohn-Sham
functional is also perfectly well defined outside this so-called
`Born-Oppenheimer surface', i.e. it is defined for any set of
orthonormal $\psi_i$'s.
   The correction terms are simple functionals of the electron density,
which can be obtained from equation (\ref{DM2rho}), and the atomic positions.
   The Kohn-Sham total energy can then be written as
\begin{eqnarray}
\label{E_KS}
 E^{KS} 
  & = & \sum_{\mu \nu} H_{\mu \nu} \rho_{\nu \mu} 
        - \frac{1}{2} \int V^H({\bf r}) 
          \rho({\bf r}) d^3{\bf r} \nonumber \\
  & + & \int (\epsilon^{xc}({\bf r}) - V^{xc}({\bf r}) ) 
       \rho({\bf r}) d^3{\bf r} \nonumber \\
  & + & \sum_{I<J} \frac{Z_I Z_J}{R_{IJ}}
\end{eqnarray}
where $I,J$ are atomic indices, 
$R_{IJ} \equiv |{\bf R}_J - {\bf R}_I|$,
$Z_I, Z_J$ are the valence ion pseudoatom charges, and 
$\epsilon^{xc}({\bf r}) \rho({\bf r})$ 
is the exchange-correlation energy density.
   In order to avoid the long range interactions of the last term,
we construct from the local-pseudopotential $V_I^{local}$,
which has an asymptotic behavior of $-Z_I/r$,
a diffuse ion charge, $\rho_I^{local}(r)$,
whose electrostatic potential is equal to $V_I^{local}(r)$:
\begin{equation}
\rho_I^{local}({\bf r}) = - \frac{1}{4 \pi} \nabla^2 V_I^{local}({\bf r}).
\end{equation}
   Notice that we define the electron density as positive, and 
therefore $\rho_I^{local} \le 0$.
   Then, we write the last term in (\ref{E_KS}) as
\begin{eqnarray}
\label{ion-ion}
   \sum_{I<J} \frac{Z_I Z_J}{R_{IJ}} 
     & = & \frac{1}{2} \sum_{IJ} U_{IJ}^{local}(R_{IJ}) 
       +   \sum_{I<J} \delta U_{IJ}^{local}(R_{IJ}) \nonumber \\
     & - & \sum_I U_I^{local}
\end{eqnarray}
where $U_{IJ}^{local}$ is the electrostatic interaction between the
diffuse ion charges in atoms $I$ and $J$:
\begin{equation}
   U_{IJ}^{local}(|{\bf R}|) = 
      \int V_I^{local}({\bf r}) \rho_J^{local}({\bf r-R}) d^3{\bf r},
\end{equation}
$\delta U_{IJ}^{local}$ is a small short-range interaction term to 
correct for a possible overlap between the soft ion charges, 
which appears when the core densities are very extended:
\begin{equation}
   \delta U_{IJ}^{local}(R) = 
      \frac{Z_I Z_J}{R} - U_{IJ}^{local}(R),
\end{equation}
and $U_I^{local}$ is the fictitious self interaction of an ion charge
(notice that the first right-hand sum in (\ref{ion-ion}) includes 
the $I=J$ terms):
\begin{equation}
   \label{U_ion}
   U_I^{local} = \frac{1}{2} U_{II}^{local}(0) = \frac{1}{2}  
                 \int V_I^{local}(r) \rho_I^{local}(r) 4 \pi r^2 dr.
\end{equation}
   Defining $\rho_I^{NA}$ from $V_I^{NA}$, analogously to 
$\rho_I^{local}$, we have that 
$\rho_I^{NA} = \rho_I^{local} + \rho_I^{atom}$, and equation 
(\ref{E_KS}) can be transformed, after some rearrangement 
of terms, into
\begin{eqnarray}
\label{E_tot}
  E^{KS}
    & = & \sum_{\mu \nu} 
          (T_{\mu \nu} + V^{KB}_{\mu \nu}) \rho_{\nu \mu} 
      +   \frac{1}{2} \sum_{IJ} U^{NA}_{IJ}(R_{IJ}) \nonumber \\
    & + & \sum_{I<J} \delta U_{IJ}^{local}(R_{IJ}) - \sum_I U_I^{local}
          \nonumber \\
    & + & \int V^{NA}({\bf r}) \delta \rho({\bf r}) d^3{\bf r} \\
    & + & \frac{1}{2} \int \delta V^H({\bf r}) \delta \rho({\bf r}) d^3{\bf r}
          + \int \epsilon^{xc}({\bf r}) \rho({\bf r}) d^3{\bf r} \nonumber
\end{eqnarray}
where $V^{NA} = \sum_I V_I^{NA}$ and
$\delta \rho = \rho - \sum_I \rho_I^{NA}$.
\begin{eqnarray}
\label{U_IJ}
   U_{IJ}^{NA}(R) 
     & = & \int V_I^{NA}({\bf r}) \rho_J^{NA}({\bf r-R}) d^3{\bf r}
           \nonumber \\
     & = & -\frac{1}{4 \pi} \int 
            V_I^{NA}({\bf r}) \nabla^2 V_J^{NA}({\bf r-R}) d^3{\bf r}
\end{eqnarray}
is a radial pairwise potential that can be obtained from $V_I^{NA}(r)$ 
as a two-center integral, by the same method described previously for
the kinetic matrix elements:
\begin{eqnarray}
\label{T_munu}
   T_{\mu \nu} 
     & = & \langle \phi_\mu | -\frac{1}{2} \nabla^2 | \phi_\nu \rangle
           \nonumber \\
     & = & -\frac{1}{2} \int 
            \phi^*_\mu({\bf r}) \nabla^2 \phi_\nu({\bf r-R}_{\mu \nu}) 
            d^3{\bf r}
\end{eqnarray}
   $V^{KB}_{\mu \nu}$ is also obtained by two-center integrals:
\begin{equation}
\label{VKB_munu}
   V^{KB}_{\mu \nu} = \sum_{\alpha} \langle \phi_\mu | \chi_\alpha \rangle
      v^{KB}_\alpha \langle \chi_\alpha | \phi_\nu \rangle
\end{equation}
where the sum is over all the KB projectors $\chi_\alpha$ that overlap 
simultaneously with $\phi_\mu$ and $\phi_\nu$.

   Although (\ref{E_tot}) is the total energy equation actually used by 
{\sc Siesta}, its meaning may be further clarified if the $I=J$ terms of 
$\frac{1}{2} \sum_{IJ} U_{IJ}^{NA}(R_{IJ})$ are
combined with $\sum_I U_I^{local}$ to yield
\begin{eqnarray}
\label{E_tot_2}
  E^{KS}
    & = & \sum_{\mu \nu} 
          (T_{\mu \nu} + V^{KB}_{\mu \nu}) \rho_{\nu \mu} 
      +   \sum_{I<J} U^{NA}_{IJ}(R_{IJ}) \nonumber \\
    & + & \sum_{I<J} \delta U_{IJ}^{local}(R_{IJ}) + \sum_I U_I^{atom}
          \nonumber \\
    & + & \int V^{NA}({\bf r}) \delta \rho({\bf r}) d^3{\bf r} \\
    & + & \frac{1}{2} \int \delta V^H({\bf r}) \delta \rho({\bf r}) d^3{\bf r}
          + \int \epsilon^{xc}({\bf r}) \rho({\bf r}) d^3{\bf r} \nonumber
\end{eqnarray}
where
\begin{equation}
\label{UI_NA}
   U_I^{atom} = \int_0^{\infty}
     \left( V_I^{local}(r) + \frac{1}{2} V_I^{atom}(r) \right)
     \rho_I^{atom}(r) 4 \pi r^2 dr
\end{equation}
is the electrostatic energy of an isolated atom.

   The last three terms in Eq.~(\ref{E_tot})
are calculated using the real space grid.
   In addition to getting rid of all long-range potentials (except that
implicit in $\delta V^H({\bf r})$), the advantage of (\ref{E_tot})
is that, apart from the relatively slowly-varying 
exchange-correlation energy density,
the grid integrals involve $\delta \rho({\bf r})$, which is generally
much smaller than $\rho({\bf r})$.
   Thus, the errors associated with the finite grid spacing are
drastically reduced. 
   Critically, the kinetic energy matrix elements can be 
calculated almost exactly, without any grid integrations.

   It is frequently desirable to introduce a finite electronic
temperature $T$ and/or a fixed chemical potential $\mu$,
either because of true physical conditions or to accelerate
the self-consistency iteration.
   Then, the functional that must be minimized is the free energy 
\cite{Mermin1965}
\begin{eqnarray}
\label{free_energy}
   F( {\bf R}_I, \psi_i({\bf r}), n_i ) 
      =  E^{KS}( {\bf R}_I, \psi_i({\bf r}), n_i )  -  \mu \sum_i n_i
              \nonumber \\
     -   k_B T \sum_i ( n_i \log n_i + (1-n_i) \log(1-n_i) ).
\end{eqnarray}
   Minimization with respect to $n_i$ yields the usual
Fermi-Dirac distribution 
$ n_i = 1/( 1 + e^{(\epsilon_i - \mu)/k_B T} ) $.

\section{Harris functional}

   We will mention here a special use of the Harris energy functional, 
that is generally defined as
\cite{Harris1985,Foulkes-Haydock1989}
\begin{eqnarray}
\label{Eharris}
   \lefteqn{
     E^{Harris}[\rho^{in}]  =  \sum_i n_i^{out}
              \langle \psi^{out}_i | \hat{H}^{in} | \psi^{out}_i \rangle }
              \nonumber \\
      && -  \frac{1}{2} \int \int 
              \frac{ \rho^{in}({\bf r}) \rho^{in}({\bf r}') }{ | {\bf r}-{\bf r}' | }
              d^3{\bf r} d^3{\bf r}'  \nonumber \\
      && +  \int ( \epsilon^{in}_{xc}({\bf r}) - 
              v^{in}_{xc}({\bf r}) ) \rho^{in}({\bf r})  d^3{\bf r}
            +  \sum_{I<J} \frac{Z_I Z_J}{R_{IJ}}
\end{eqnarray}
where $\hat{H}^{in}$ is the KS Hamiltonian produced by a trial density 
$\rho^{in}$ and
$\psi^{out}_i$ are its eigenvectors (which in general are different from
those whose density is $\rho^{in}$).
   As in Eq.~(\ref{E_BS}), the first term in ({\ref{Eharris}) can be written
as $\mbox{Tr}( H^{in} \rho^{out} )$, and the rest are the so-called
`double count corrections'.
   An important advantage of eq. (\ref{Eharris}) is that it does not require
$\rho^{in}_i$ to be obtained from a set of orthogonal electron states
$\psi^{in}_i$, and in fact  $\rho^{in}$ is frequently taken as a simple
superposition of atomic densities.
   However, we will assume here that the states $\psi^{in}_i$ are indeed
known.
   In this case, the Kohn-Sham energy $E^{KS}[\rho^{in}]$,
Eq.~(\ref{E_KS}),  obeys exactly the same expression (\ref{Eharris}), 
except that $\psi^{out}_i$ and $n_i^{out}$ must be replaced by 
$\psi^{in}_i$ and $n_i^{in}$.
   Thus, a simple subtraction gives
\begin{equation}
\label{DEharris}
   E^{Harris}[\rho^{in}] = E^{KS}[\rho^{in}] + 
     \sum_{\mu\nu} H^{in}_{\nu\mu} 
     \left( \rho^{out}_{\mu\nu} - \rho^{in}_{\mu\nu} \right).
\end{equation}
   Generally the Harris functional is used nonself-consistently,
with a trial density given by the sum of atomic densities.
   But here we want to comment on its usefulness to improve
dramatically the estimate of the converged total energy,
by taking $\rho^{in}_{\mu\nu}$ as the density matrix of
the $(n-1)$'th self-consistency
iteration and $\rho^{out}_{\mu\nu}$ of the $n$'th iteration.
   In fact, $E^{Harris}$ frequently gives, after just two or three 
iterations, a better estimate than $E^{KS}$ after tens of iterations.
   Unfortunately, we have found that there is hardly any improvement
in the convergence of the atomic forces thus estimated, and 
therefore the self-consistent Harris functional is less 
useful for geometry relaxations or molecular dynamics.

\section{Atomic forces}

   Atomic forces and stresses are obtained by direct differentiation of 
(\ref{E_tot}) with respect to atomic positions.
   They are obtained simultaneously with the total energy,
mostly in the same places of the code, under the general paradigm
``a piece of energy $\Rightarrow$ a piece of force/stress''
(except that some pieces are calculated only in the last
self-consistency step).
   This ensures that all force contributions, including Pulay 
corrections, are automatically included.
   The force contribution from the first term in (\ref{E_tot}) is
\begin{eqnarray}
\label{dEkdr}
   \frac{ \partial }{ \partial {\bf R}_I }
   &&    \sum_{\mu \nu} 
         (T_{\mu \nu} + V^{KB}_{\mu \nu}) \rho_{\nu \mu} =  
         \nonumber \\
   &&    \sum_{\mu \nu} 
         ( T_{\mu \nu} + V^{KB}_{\mu \nu} )
         \frac{ \partial \rho_{\nu \mu} }{ \partial {\bf R}_I }
       + 2 \sum_\mu \sum_{\nu \in I} 
         \frac{ d T_{\mu \nu} }{ d {\bf R}_{\mu \nu} } \rho_{\nu \mu}
         \nonumber \\
   &&  + 2 \sum_\mu \sum_{\nu \in I} \sum_\alpha
         S_{\mu \alpha} v^{KB}_\alpha 
         \frac{ d S_{\alpha \nu} }{ d {\bf R}_{\alpha \nu} }  
         \rho_{\nu \mu} \nonumber \\
   &&  - 2 \sum_{\mu \nu} \sum_{\alpha \in I} 
          S_{\mu \alpha} v^{KB}_\alpha  
         \frac{ d S_{\alpha \nu} }{ d {\bf R}_{\alpha \nu} }
         \rho_{\nu \mu}
\end{eqnarray}
where $\alpha$ are KB projector indices,
$\in I$ indicates orbitals or KB projectors belonging to atom $I$,
and we have considered that
\begin{equation}
\label{drhoTdrI}
     \frac{\partial S_{\mu \nu}}{\partial {\bf R}_{I_\nu}} =
   - \frac{\partial S_{\mu \nu}}{\partial {\bf R}_{I_\mu}} =
     \frac{d S_{\mu \nu}}
          {d {\bf R}_{\mu \nu}},
\end{equation}
where ${\bf R}_{I_\mu}$ is the position of atom $I_\mu$, to which
orbital $\phi_\mu$ belongs and 
${\bf R}_{\mu \nu} = {\bf R}_{I_\nu} - {\bf R}_{I_\mu}$.

   Leaving aside for appendix \ref{Orthogonality_force} the terms 
containing $\partial \rho_{\nu \mu} / \partial {\bf R}_I$, the other 
derivatives can be obtained by straightforward differentiation of 
their expansion in spherical harmonics (Eq.~(\ref{S_final})).
   However, instead of using the spherical harmonics 
$Y_{lm}(\hat{\bf r})$ themselves, it is convenient to multiply them
by $r^l$, in order to make them analytic at the origin.
   Thus
\begin{eqnarray}
\label{dTmunudr}
   \frac{d S_{\mu \nu}({\bf R})}{d {\bf R}} 
   & = & \sum_{lm} \nabla \left( \frac{S^{\mu \nu}_{lm}(R)}{R^l}
         R^l Y_{lm}(\hat{\bf R}) \right) \nonumber \\
   & = & \sum_{lm}
         \frac{d}{dR} \left( \frac{S^{\mu \nu}_{lm}(R)}{R^l} \right)
         R^l Y_{lm}(\hat{\bf R}) \hat{\bf R} \nonumber \\
   & + & \sum_{lm} \frac{S^{\mu \nu}_{lm}(R)}{R^l}
         \nabla ( R^l Y_{lm}(\hat{\bf R}) )
\end{eqnarray}
   In fact, it is $S^{\mu \nu}_{lm}(R)/R^l$, rather than 
$S^{\mu \nu}_{lm}(R)$, that is stored as a function of $R$ on a 
radial grid.
   Its derivative, $d(S^{\mu \nu}_{lm}(R)/R^l)/dR$, is then obtained 
from the same cubic spline interpolation used for the value itself.
   The value and gradient of $R^l Y_{lm}(\hat{\bf R})$ are calculated
analytically from explicit formulae (up to $l=2$) or recurrence
relations
 \cite{NumericalRecipes}.
   Entirely analogous equations apply to 
$dT_{\mu \nu} / d{\bf R}_{\mu \nu}$.

   The second and third terms in Eq.~(\ref{E_tot})
are simple interatomic pair potentials whose force contributions are
calculated trivially from their radial spline interpolations.
   The fourth term is a constant which does not depend on the atomic 
positions.
   Taking into account that 
$V^{NA}({\bf r}) = \sum_I V^{NA}_I({\bf r-R}_I)$,
and therefore 
$\partial V^{NA}({\bf r}) / \partial{\bf R}_I = 
  - \nabla V^{NA}_I({\bf r-R}_I)$,
the force contribution from the fifth term is
\begin{eqnarray}
\label{dENAdr}
   \lefteqn{
    \frac{ \partial }{ \partial {\bf R}_I }
         \int V^{NA}({\bf r}) \delta \rho({\bf r}) d^3{\bf r} = } \\
    && - \int \nabla V^{NA}_I({\bf r}) \delta \rho({\bf r}) d^3{\bf r}
         + \int V^{NA}({\bf r}) 
          \frac{ \partial \delta \rho({\bf r}) }{ \partial {\bf R}_I}
          d^3{\bf r} \nonumber
\end{eqnarray}
   The sixth term is the electrostatic self-energy of the charge
distribution $\delta \rho({\bf r})$:
\begin{equation}
\label{dEHdr}
   \frac{\partial}{\partial {\bf R}_I} \frac{1}{2}
         \int \delta V^H({\bf r}) \delta \rho({\bf r}) d^3{\bf r} 
     = \int \delta V^H({\bf r}) 
       \frac{\partial \delta \rho({\bf r})}{\partial {\bf R}_I} d^3{\bf r}
\end{equation}
   In the last term, we take into account that 
$d(\rho \epsilon^{xc}) / d \rho = v^{xc}$ to obtain
\begin{equation}
\label{dExcdr}
   \frac{\partial}{\partial {\bf R}_I} 
         \int \epsilon^{xc}({\bf r}) \rho({\bf r}) d^3{\bf r} 
      =  \int V^{xc}({\bf r}) 
         \frac{\partial \rho({\bf r})}{\partial {\bf R}_I} d^3{\bf r}
\end{equation}
   Now, using Eq.~(\ref{DM2rho}) and that, for $\nu \in I$,
$\partial \phi_\nu({\bf r}) / \partial {\bf R}_I = - \nabla \phi_\nu$,
the change of the self-consistent and atomic densities are
\begin{eqnarray}
\label{drhodr}
   \frac{ \partial \rho({\bf r}) }{ \partial {\bf R}_I } 
    & = & \mbox{Re}  \sum_{\mu \nu}
               \frac{ \partial \rho_{\nu \mu} }{ \partial {\bf R}_I }
               \phi^*_\mu({\bf r}) \phi_\nu({\bf r}) \nonumber \\
    & - & 2 \mbox{Re} \sum_{\mu} \sum_{\nu \in I}
                \rho_{\nu \mu} \phi^*_\mu({\bf r}) 
                \nabla \phi_{\nu}({\bf r})
\end{eqnarray}
\begin{equation}
\label{drho0dr}
   \frac{ \partial \rho^{atom}({\bf r}) }{ \partial {\bf R}_I } =
      - 2  \mbox{Re} \sum_{\mu \in I} \rho^{atom}_{\mu \mu} 
            \phi^*_\mu({\bf r}) \nabla \phi_{\mu}({\bf r})
\end{equation}
where we have taken into account that the density matrix of the
separated atoms is diagonal.
   Thus, leaving still aside the terms with
$\partial \rho_{\nu \mu} / \partial {\bf R}_I$, 
the last term in Eq.~(\ref{dENAdr}),
as well as those in (\ref{dEHdr}) and (\ref{dExcdr}), 
have the general form
\begin{eqnarray}
\label{dEVdr}
\lefteqn{
    \mbox{Re} \sum_{\mu} \sum_{\nu \in I} \rho_{\nu\mu} 
         \int V({\bf r}) \phi^*_\mu({\bf r}) \nabla \phi_\nu({\bf r})
            d^3{\bf r} } \nonumber \\
    && =  \mbox{Re} \sum_{\mu}  \sum_{\nu \in I} \rho_{\nu\mu}
          \langle \phi_\mu | V({\bf r}) | \nabla \phi_\nu \rangle.
\end{eqnarray}
   These integrals are calculated on the grid, in the same way
as those for the total energy
(i.\ e.\ $\langle \phi_\mu | V({\bf r}) | \phi_\nu \rangle$).
   The gradients $\nabla \phi_\nu({\bf r})$ at the grid points
are obtained analytically, like those of $\phi_\nu({\bf r})$  
from their radial grid interpolations of $\phi(r) / r^l$:
\begin{eqnarray}
\label{dphidr}
   \nabla \phi_{Ilmn}({\bf r}) 
   & = & \frac{d}{dr} 
         \left( \frac{ \phi_{Iln}(r) }{ r^l } \right)
         r^l Y_{lm}(\hat{\bf r}) \hat{\bf r} \nonumber \\
   & + & \frac{\phi_{Iln}(r)}{r^l}
         \nabla ( r^l Y_{lm}(\hat{\bf r}) ).
\end{eqnarray}

   In some special cases, with elements that require 
hard partial core corrections or explicit inclusion of the semicore, 
the grid integrals may pose a problem for geometry relaxations, 
because they make the energy dependent on the position of the 
atoms relative to the grid.
   This `eggbox effect' is small for the energy itself, and it 
decreases fast with the grid spacing.
   But  the effect is larger and the convergence slower for the 
forces, as they are proportional to the amplitude of the energy 
oscillation, but inversely proportional to its period.
   These force oscillations  complicate the force
landscape, especially when the true atomic forces become small,
making the convergence of the geometry optimization more difficult.
   Of course, the problem can be avoided by decreasing the grid 
spacing but this has an additional cost in computer time and memory.
   Therefore, we have found it useful to minimize this problem
by recalculating the forces, at a set of positions, determined 
by translating the whole system by a set of points in a finer mesh.
   This procedure, which we call `grid-cell sampling',
has no extra cost in memory.
   And since it is done only at the end of the self-consistency 
iteration, for fixed $\rho_ {\mu \nu}$, it has only a moderate 
cost in CPU time.

   At finite temperature, the forces are really the derivatives of 
the {\em free} energy with respect to atomic displacements since
\begin{eqnarray}
\label{force_finiteT}
\lefteqn{
   \frac{dF({\bf R}_I,\psi_i({\bf r}),n_i)}{d{\bf R}_I}
       = \frac{\partial F}{\partial {\bf R}_I}
        + \sum_i \frac{\partial F}{\partial n_i}
                     \frac{\partial n_i}{\partial {\bf R}_I} } \nonumber \\
     && + \sum_i \int \frac{\partial F}{\partial \psi^*_i({\bf r})}
                              \frac{\partial \psi_i({\bf r})}{\partial  {\bf R}_I}
                         d^3 {\bf r}
         = \frac{\partial E}{\partial {\bf R}_I}.
\end{eqnarray}
   In this particular equation we have used the notation $d/d{\bf R}_I$, 
as opposed to $\partial / \partial {\bf R}_I$, to indicate the
inclusion of the change in $\psi_i({\bf r})$ and $n_i$ when we 
move the atom, in calculating the derivative.
   But we have used also that 
$\partial F/ \partial n_i = \partial F / \partial \psi_i({\bf r}) = 0$
and that the last two terms in (\ref{free_energy}) do not depend
on ${\bf R}_I$, so that 
$\partial F/ \partial {\bf R}_I = \partial E / \partial {\bf R}_I$.
   The latter are the atomic forces actually calculated.
   Notice, however, that 
$dE/d{\bf R}_I \ne \partial E / \partial {\bf R}_I$, so that
the calculated forces are indeed the total derivatives of
the free, not the internal energy.

   We would like to also mention the calculation of forces
using the non-self-consistent Harris functional, in which the
`in' density is a superposition of atomic densities.
   We have implemented this as an option for `quick and dirty'
calculations because, used with a minimal basis set, it makes
{\sc Siesta} competitive with tight binding methods, which are 
much faster than density functional calculations.
   The problem that we address here is that, although $E^{Harris}$
is stationary with respect to $\rho^{out}$, it is not so with
respect to $\rho^{in}$.
   In particular, there appears a force term
\begin{equation}
\label{F_Harris}
   \int \frac{ \partial V_{xc}^{in}({\bf r}) }{ \partial {\bf R}_I }
   \rho^{out}({\bf r}) d^3{\bf r}.
\end{equation}
   A similar term appears for the electrostatic interaction 
between the input and output density, but it presents no special
problems because of the linear character of the Hartree potential.
   However, evaluation of (\ref{F_Harris}) requires the change
of the exchange-correlation potential with density, a quantity
also required to evaluate the linear response of the electron gas,
but not in normal energy and force calculations.
   Finally, notice that, apart from this minor difficulty, the 
Harris-functional forces are perfectly well defined {\em at the 
first iteration only}.
   For later iterations (but still not converged) there is no
practical way to calculate $\partial \rho^{in} / \partial {\bf R_I}$ and,
without the help of Hellman-Feynmann's theorem
(which applies only at convergence), the forces are
not well defined.
   Of course, the omission of the terms depending on this 
quantity produces an estimate of the forces, but we have found
that their convergence is not appreciably faster than those
estimated from the Kohn-Sham functional.

\section{Stress tensor}

   We define the stress tensor as the positive derivative of the 
total energy with respect to the strain tensor
\begin{equation}
\label{stress}
   \sigma_{\alpha\beta} = 
     \frac{ \partial E^{KS} }{ \partial \epsilon_{\alpha\beta} }
\end{equation}
where $\alpha,\beta$ are Cartesian coordinate indices.
   To translate to standard units of pressure, we must simply
divide by the unit-cell volume and change sign.
   During the deformation, all vector positions, including those
of atoms and grid points (and of course lattice vectors), change
according to
\begin{equation}
\label{strain}
   r'_\alpha = \sum_{\beta=1}^3 
      \left( \delta_{\alpha\beta} + \epsilon_{\alpha\beta} \right)
      r_\beta
\end{equation}
   The shape of the basis functions, KB projectors, and atomic
densities and potentials do not change, but their origin gets
displaced according to (\ref{strain}).
   From this equation, we find that
\begin{equation}
\label{drdeps}
   \frac{ \partial r_\gamma }{ \partial \epsilon_{\alpha\beta} } =
   \delta_{\gamma\alpha} r_\beta
\end{equation}
   The change in $E^{KS}$ is essentially due to these position 
displacements, and therefore the calculation of the stress is
almost perfectly parallel to that of the atomic forces,
thus being performed in the same sections of the code.
   For example:
\begin{equation}
\label{dTdeps}
   \frac{ \partial T_{\mu\nu} }{ \partial \epsilon_{\alpha\beta} } 
      = \sum_{\gamma=1}^3 
        \frac{ \partial T_{\mu\nu} }{ \partial r^\gamma_{\mu\nu} }
        \frac{ \partial r^\gamma_{\mu\nu} }
             { \partial \epsilon_{\alpha\beta} }
      = \frac{ \partial T_{\mu\nu} }{ \partial r^\alpha_{\mu\nu} }
         r^\beta_{\mu\nu}
\end{equation}
   Since $\partial T_{\mu\nu} / \partial r^\alpha_{\mu\nu}$ is
evaluated to calculate the forces, it takes very little extra
effort to multiply it also by $r^\beta_{\mu\nu}$ for the stress.
   Equally, force contributions like (\ref{dEVdr}) have their
obvious stress counterpart
\begin{equation}
\label{dEVdeps}
    \sum_{\mu\nu} \rho_{\nu\mu} 
       \langle \phi_\mu | V({\bf r}) | 
       (\nabla_\alpha \phi_\nu) r_\beta \rangle
\end{equation}

   However, there are three exceptions to this parallelism.
   The first concerns the change of the volume per grid point or,
in other words, the Jacobian of the transformation (\ref{strain})
in the integrals over the unit cell.
   This Jacobian is simply $\delta_{\alpha\beta}$, and it leads to
a stress contribution
\begin{equation}
\label{stress_jacobian}
   \left[ \int \left( V^{NA}({\bf r}) + \frac{1}{2} \delta V^H({\bf r}) \right)
        \delta \rho({\bf r}) d^3{\bf r} + E^{xc} \right]
   \delta_{\alpha\beta}
\end{equation}
   Notice that the the renormalization of the density, required to 
conserve the charge when the volume changes, enters through 
the orthonormality constraints, to be discussed in appendix
\ref{Orthogonality_force}.
   The second special contribution to the stress lies in the fact
that, as we deform the lattice, there is a change in the factor
$1/|{\bf r - r}'|$ of the electrostatic energy integrals.
   We deal with this contribution in reciprocal space, when we
calculate the Hartree potential by FFTs, by evaluating the 
derivative of the reciprocal-space vectors with respect to
$\epsilon_{\alpha\beta}$.
   Since 
$G'_\alpha = \sum_\beta G_\beta 
        ( \delta_{\beta \alpha} - \epsilon_{\beta \alpha} )$:
\begin{equation}
\label{dEHdeps}
     \frac{ \partial }{ \partial \epsilon_{\alpha\beta} } 
      \frac{ 1 }{ G^2 } 
    = \frac{ 2 G_\alpha G_\beta }{ G^4 } 
\end{equation}
   Finally, the third special stress contribution arises in GGA 
exchange and correlation, from the change of the gradient of the 
deformed density $\rho({\bf r}) \rightarrow \rho({\bf r}')$.
   The treatment of this contribution is explained in detail in
reference \onlinecite{Balbas2001}.

\section*{Electric Polarization }

   The calculation of the electric polarization, as an integral 
in the grid across the unit cell, is standard and almost 
free for molecules, chains and slabs (in the directions
perpendicular to the chain axis, or to the surface). 
   For bulk systems, the electric polarization cannot be found from 
the charge distribution in the unit cell alone.
   In this case, we need the so-called Berry-phase theory
of polarization~\cite{KingSmith-Vanderbilt1993,Resta1994},
which allows to compute
quantities like the dynamical charges~\cite{KingSmith-Vanderbilt1993} 
and piezoeletric constants~\cite{SaghiSzabo1998,Vanderbilt2000}.
   Here we comment some details of 
our implementation \cite{SanchezPortal2000}.

   If ${\bf R}_\alpha$ are the lattice vetors and 
${\bf P}^e= \sum_{\alpha=1}^3 P_\alpha^e {\bf R}_\alpha$ 
is the electronic contribution to the macroscopic polarization, 
then we have  
\begin{eqnarray}
\label{pol_eq}
     2 \pi \; P_{\alpha}^e 
  & = &  {\bf G}_\alpha \cdot {\bf P}^e \nonumber \\
  & = &  - \frac{2 e}{(2 \pi)^3} 
            \int_{_{BZ}} d{\bf k} \;\;
            {\bf G}_\alpha \cdot \left. \frac{\partial}{\partial {\bf k}'} 
        \Phi({\bf k},{\bf k}^\prime)  \right|_{{\bf k}^\prime={\bf k}}
\end{eqnarray}
where ${\bf G}_\alpha$ is the corresponding reciprocal lattice 
vector, $e$ is the electron charge, 
$u_{i}({\bf k},{\bf r})=e^{-i{\bf k.r}} \psi_{i}({\bf k},{\bf r})$ 
is the periodic part of the Bloch function, and the factor 
of two comes from the spin degeneracy.
   The quantum phase $\Phi({\bf k},{\bf k}^\prime)$ is defined as 
\begin{equation}
\label{phase_eq}
       \Phi({\bf k},{\bf k}^\prime)=
\text{Im} \left[ \text{ln} \left( \text{det}            
\langle u_{i}({\bf k},{\bf r})|
u_{j}({\bf k^\prime},{\bf r}) \rangle 
\right) \right]
\end{equation}
   The derivative  in (\ref{pol_eq}) depends on a  gauge
that must be chosen such that 
$u({\bf k+G},{\bf r})=e^{-i {\bf G \cdot r}}
u({\bf k},{\bf r})$. 
   In practice, the integral is replaced by a discrete
summation, and a finite-difference approximation is made for 
the derivative~\cite{KingSmith-Vanderbilt1993}: 
$\Delta {\bf k}_\alpha \left. \frac{\partial}{\partial  {\bf k}'_\alpha} 
 \Phi({\bf k},{\bf k}^\prime)
\right|_{{\bf k}^\prime={\bf k}} \approx \frac{1}{2}
[ \Phi({\bf k},{\bf k}+\Delta {\bf k}_\alpha) - 
\Phi({\bf k},{\bf k}-\Delta  {\bf k}_\alpha)]$,
where $\Delta  {\bf k}_\alpha = {\bf G}_\alpha / N_\alpha$.
   Then  (\ref{pol_eq}) becomes, for $\alpha=1$ 
\begin{equation}
\label{pol_discret}
     {\bf G}_{1} \cdot {\bf P}_e \approx
    - \frac{2 e}{\Omega N_2 N_3}
      \sum_{i_2=0,i_3=0}^{N_2-1,N_3-1} 
   \sum_{i_1=0}^{N_1-1}
  \Phi({\bf k}_{i_1 i_2 i_3},{\bf k}_{i_1+1 i_2 i_3}) \; ,
\end{equation}
where we have split the sum to stress the fact
that we have a two-dimensional integral in the plane defined
by ${\bf G}_2$ and ${\bf G}_3$, and a linear integral along
${\bf G}_1$. Due to the approximation in the derivative, the 
linear integral usually requires a finer mesh
than the surface integral.
   To evaluate $\Phi({\bf k},{\bf k}+\Delta {\bf k})$ 
we use our LCAO basis:
\begin{eqnarray}
\label{phase_eq_2}
&
\langle u_{i}({\bf k})|
u_{j}({\bf k}+\Delta {\bf k}) \rangle = 
\langle \psi_{i}({\bf k})| 
e^{-i \Delta {\bf k} \cdot {\bf r}}|
\psi_{j}({\bf k}+\Delta {\bf k}) \rangle = &
\nonumber \\
&
\sum_{\nu} \sum_{\mu{^\prime}} 
c_{i \nu}({\bf k}) c_{\mu' j}({\bf k}+\Delta {\bf k}) &
\\
&
e^{-i{\bf k} \cdot ({\bf R}_{\nu}-{\bf R}_{\mu^{\prime} }) } 
\langle \phi_\nu |
 e^{-i \Delta {\bf k} \cdot ({\bf r}-{\bf R}_{\mu^{\prime}})}|
\phi_{\mu^{\prime}} \rangle . \nonumber
\end{eqnarray}
   Formulas similar to (\ref{phase_eq_2}) have been implemented 
by several authors~\cite{Dallolio1997,Yaschenko1998}, mainly
in the context of Hatree-Fock calculations, in which
the basis orbitals are expanded in gaussians whose matrix 
elements can be found analytically~\cite{Dallolio1997}. 
   Our numerical, localized pseudo-atomic basis
orbitals are not well suited for a gaussian expansion.
   Instead, we expand the plane-waves appearing in equation 
(\ref{phase_eq_2}) to first order in $\Delta {\bf k}$,
$e^{-i \Delta {\bf k} \cdot ({\bf r}-{\bf R}_\mu)}\approx
1-i \Delta {\bf k} \cdot ({\bf r}-{\bf R}_\mu)
+{\cal O}(\Delta k^2)$, and then we calculate the matrix
elements of the position operator as explained in 
section\ref{Two-center}.
   It is interesting to note that, since the
discretized formula (\ref{pol_discret}) only holds to 
${\cal O}(\Delta k^2)$,
the approximation of the matrix elements in
(\ref{phase_eq_2}) does not introduce
any further errors in the calculation of the polarization.
   In  a symmetrized version, we approximate
equation (\ref{phase_eq_2}) as
\begin{eqnarray}
\label{position_eq}
\sum_{\nu} \sum_{\mu{^\prime}}
   c_{i \nu}({\bf k}) c_{\mu' j}({\bf k} + \Delta {\bf k})
   e^{-i({\bf k}+\frac{\Delta {\bf k}}{2}) 
\cdot ({\bf R}_{\nu}-{\bf R}_{\mu^{\prime} }) }
\left[ \;
\langle \phi_\nu |\phi_{\mu^{\prime}} \rangle 
\right.
\nonumber \\
    -i \frac{\Delta {\bf k}}{2} \cdot \left( \left.
       \langle \phi_\nu | ({\bf r}-{\bf R}_{\nu})   | \phi_{\mu'} \rangle
    + \langle \phi_\nu | ({\bf r}-{\bf R}_{\mu' }) | \phi_{\mu'} \rangle
\right) \;
\right],
\end{eqnarray}

\section{Order-$N$ functional}

   The basic problem for solving the Kohn-Sham equations in 
${\cal O}(N)$ operations is that the solutions (the Hamiltonian
eigenvectors) are extended over the whole system and overlap
with each other.
   Just to check the orthogonality of $N$ trial solutions, by
performing integrals over the whole system,  involves 
$\sim N^3$ operations.
   Among the different methods proposed to solve this problem
\cite{Ordejon1998:revTBON,Goedecker1999:RMP},
we have chosen the localized-orbital approach
\cite{Mauri-Galli-Car1993,Ordejon1993,Kim-Mauri-Galli1995}
because of its superior efficiency for non-orthogonal basis sets.
   The initially proposed functional
\cite{Mauri-Galli-Car1993,Ordejon1993}
used a fixed number of occupied states, equal to the number of 
electron pairs, and it was found to have numerous local minima
in which the electron configuration was easily trapped.
   A revised functional form
\cite{Kim-Mauri-Galli1995}
which uses a larger number of states than electron pairs, 
with variable occupations, has been found empirically to avoid
the local minima problem.
   This is the functional that we use and recommend.

   Each of the localized, Wannier-like states, is constrained
to its own localization region.
   Each atom $I$ is assigned a number of states equal to 
int($Z_I^{val}$/2+1) so that, if doubly occupied,  they can 
contain at least one excess electron (they can also become 
empty during the minimization of the energy functional).
   These states are  confined to a sphere of radius $R_c$ 
(common to all states) centered at ${\bf R}_I$.
   More precisely, the expansion (Eq.~(\ref{psi})) of a state  
$\psi_i$ centered at ${\bf R}_I$  may contain only basis orbitals 
$\phi_\mu$ centered on atoms $J$ such that $|{\bf R}_{IJ}| < R_c$.
   This implies that $\psi_i({\bf r})$ may extend to a maximum
range $R_c + r_c^{max}$, where $r_c^{max}$ is the maximun
range of the basis orbitals.
   For covalent systems, a localization region centered on 
bonds rather than atoms is more efficient \cite{Stephan1998}
(it leads to a lower energy for the same $R_c$), but it is less 
suitable to a general algorithm, especially in case of ambiguous 
bonds.
   Therefore, we generally use the atom-centered localization 
regions.

   In the method of Kim, Mauri, and Galli (KMG)
\cite{Kim-Mauri-Galli1995}, 
the band-structure energy is rewritten as:
\begin{eqnarray}
\label{E_KMG}
  E^{KMG}
    & = & 2 \sum_{ij} (2 \delta_{ji} - S_{ji} ) ( H_{ij} - \eta S_{ij} )
             \nonumber \\
    & = & 4 \sum_{i} \sum_{\mu \nu}
                 c_{i \mu} \delta H_{\mu \nu} c_{\nu i}
             \nonumber \\
    & - & 2 \sum_{ij} \sum_{\alpha \beta \mu \nu}
                c_{i \alpha} S_{\alpha \beta} c_{\beta j}
                c_{j \mu} \delta H_{\mu \nu} c_{\nu i}
\end{eqnarray}
    Where 
$S_{ij} = \langle \psi_i | \psi_j \rangle$,
$H_{ij} = \langle \psi_i | H | \psi_j \rangle$,
$\delta H_{\mu \nu} = H_{\mu \nu} - \eta S_{\mu \nu}$,
and we have assumed a non-magnetic solution with doubly occupied
states.
    The `double count' correction terms of Eq.~(\ref{E_KS}) remain 
unchanged and the electron density is still defined by (\ref{DM2rho}),
but the density matrix is re-defined as
\begin{eqnarray}
\label{rho_KMG}
   \rho_{\mu \nu} 
     & = & 2 \sum_{ij}  c_{\mu i} (2 \delta_{ij} - S_{ij} ) c_{j \nu}
              \nonumber \\
     & = & 4 \sum_{i} c_{\mu i} c_{i \nu}
        -   2  \sum_{ij} \sum_{\alpha \beta}
               c_{\mu i} c_{i \alpha} S_{\alpha \beta} c_{\beta j} c_{j \nu}
\end{eqnarray}

    The parameter $\eta$ in Eq.~(\ref{E_KMG}) plays the role of a
chemical potential, and must be chosen to lie within the band gap
between the occupied and empty states.
   This may be tricky sometimes, since the electron bands can shift
during the self-consistency process or when the atoms move.
   In general, the number of electrons will not be exactly the
desired one, even if $\eta$ is within the band gap, because
the minimization of (\ref{E_KMG}) implies a trade-off in which 
the localized states become fractionally occupied.
   To avoid an infinite Hartree energy in periodic systems, 
we simply renormalize the density matrix so that the total 
electron charge $\sum_{\mu \nu} S_{\mu \nu} \rho_{\nu \mu} $
equals the required value.

   For a given potential, the functional (\ref{E_KMG}) is
minimized by the conjugate-gradients method, using its
derivatives with respect to the expansion coefficients
\begin{eqnarray}
\label{gradE_KMG}
   \frac{ \partial E^{KMG} }{ \partial c_{i \mu} }
    & = & 4 \sum_{\nu} \delta H_{\mu \nu} c_{\nu i} 
              \nonumber \\
       -  2 \sum_j \sum_{\alpha \beta \nu}
   & (  &  S_{\mu \nu} c_{\nu j} c_{j \alpha} 
                     \delta H_{\alpha \beta} c_{\beta i}
              \nonumber \\
   & + &  \delta H_{\mu \nu}  c_{\nu j} c_{j \alpha} 
                     S_{\alpha \beta} c_{\beta i} )
\end{eqnarray}
   The minimization proceeds without need to orthonormalize
the electron states $\psi_i$.
   Instead, the orthogonality, as well as the correct normalization
(one below $\eta$ and zero above it) result as a consequence
of the minimization of $E^{KMG}$.
   This is because, in contrast to the KS functional,  $E^{KMG}$
is designed to penalize any nonorthogonality
\cite{Kim-Mauri-Galli1995}.
   The KS ground state, with all the occupied $\psi_i$'s 
orthonormal, is also the minimum of (\ref{E_KMG}), at which
$E^{KMG}=E^{KS}$.
   If the variational freedom is constrained by the localization
of the $\psi_i$'s, the orthogonality cannot be exact, and the
resulting energy is slightly larger than for unconstrained
wavefunctions.
   In insulators and semiconductors, the Wannier functions
are exponentially localized \cite{Kohn1959}, and the
energy excess due to their strict localization
decreases rapidly as a function of the localization radius 
$R_c$, as can be seen in Fig.~\ref{RcConv}.

\begin{figure}[htbp]
\includegraphics[width=\columnwidth] {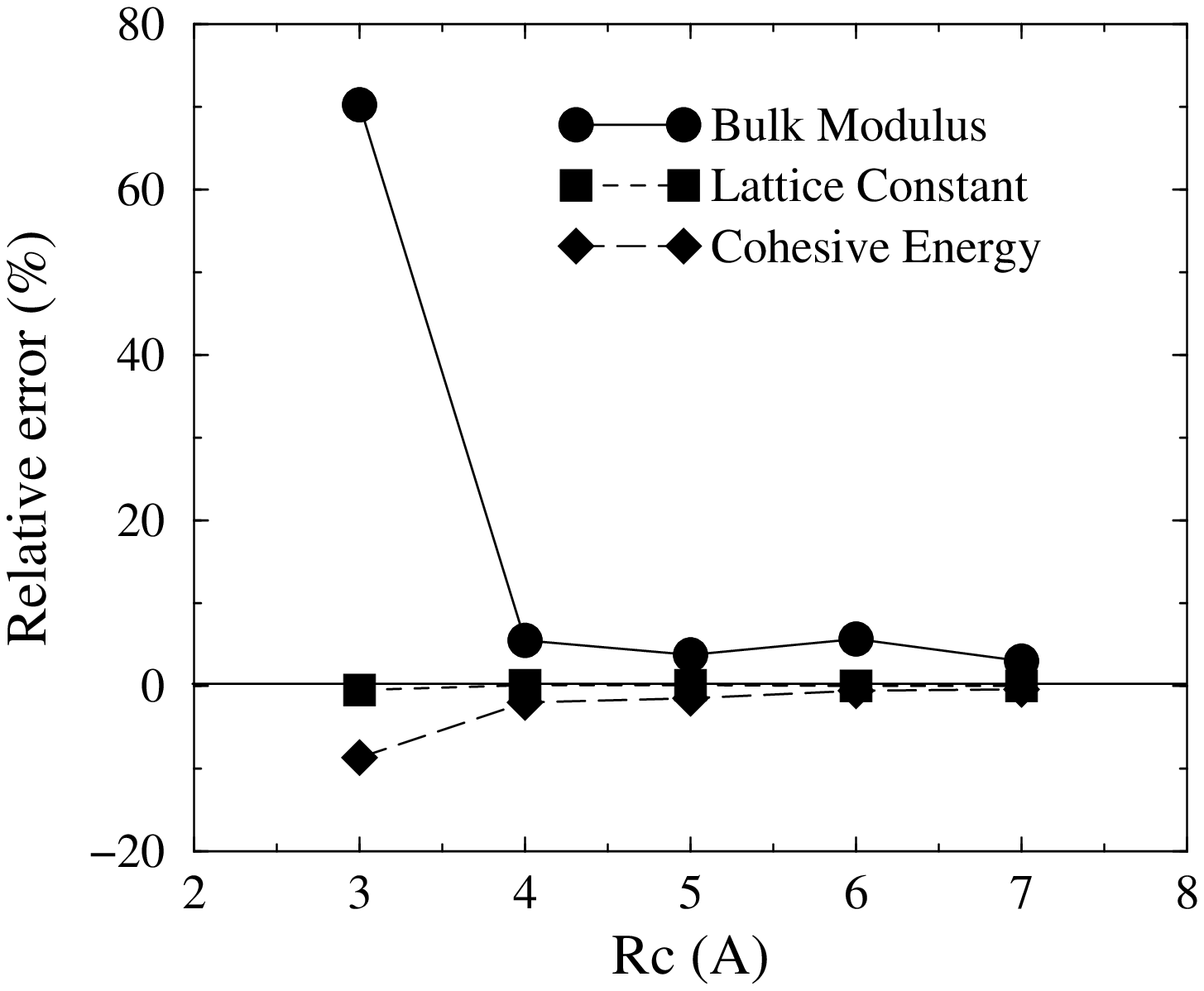}
   \caption{   Convergence of the lattice constant,
                bulk modulus, and cohesive energy
                as a function of the localization radius $R_c$ of
                the Wannier-like electron states in silicon.
                   We used a supercell of 512 atoms and a minimal 
                basis set with a cutoff radius $r_c = 5$~a.u.\ for 
                both $s$ and $p$ orbitals.
               }
\label{RcConv}
\end{figure}

   If the system is metallic, or if the chemical potential is not
within the band gap (for example because of the presence of
defects), the KMG functional cannot be used in practice.
   In fact, although some ${\cal O}(N)$ methods can handle
metallic systems in principle \cite{Goedecker1999:RMP},
we are not aware of any practical calculations at a DFT level.
   In such cases we copy the Hamiltonian and overlap matrices
to standard expanded arrays and solve the generalized 
eigenvalue problem by conventional order-$N^3$ diagonalization 
techniques  \cite{LAPACK1999}.
   However, even in this case, most of the operations, and 
particularly those to find the density and potential, and to set 
up the Hamiltonian, are still performed in ${\cal O}(N)$ operations.

   Irrespective of whether the ${\cal O}(N)$ functional or the 
standard diagonalization is used, an outer self-consistency
iteration is required, in which the density matrix 
is updated using Pulay's  Residual Metric Minimization by 
Direct Inversion of the Iterative Subspace (RMM-DIIS) method
\cite{Pulay1980,Pulay1982}.
   Even when the code is strictly ${\cal O}(N)$, the CPU time may
increase faster if the number of iterations required to achieve the
solution increases with $N$.
   In fact, it is a common experience that the required number of  
selfconsistency iterations increases with the size of the system.
   This is mainly because of the `charge sloshing' effect, 
in which small displacements of charge from one side of
the system to another give rise to larger changes 
of the potential, as the size increases.
   Fortunately, the localized character of the Wannier-like
wavefunctions used in the ${\cal O}(N)$ method help to solve
also this problem, by limiting  the charge sloshing.
   Table~\ref{iterations} presents the average number of iterations 
required to minimize the ${\cal O}(N)$ functional  
and the average number of selfconsistency iterations,
during a molecular dynamics simulation of bulk silicon at
room temperature.
   It can be seen that these numbers are quite small and that
they increase very moderately with system size.
   As might be expected, the number of minimization iterations
increase with the localization radius, i.e.\ with the number of
degrees of freedom ($c_{\mu i}$ coefficients) of the wavefunctions.
   But this increase is also rather moderate.

\begin{table}
\caption[ ]{
   Average number of selfconsistency (SCF) iterations (per molecular
dynamics step) and average number of conjugate-gradient (CG) 
iterations (per SCF iteration) required to minimize the  ${\cal O}(N)$ 
functional, during a simulation of bulk silicon at $\sim 300$ K.
   We used the Verlet method~\cite{Allen-Tildesley1987}
at constant energy, with a time step of 1.5 fs, and
a  minimal basis set with a cutoff radius $r_c = 5$ a.u.
   $R_c$ is the localization radius of the Wannier-like wavefunctions
used in the  ${\cal O}(N)$ functional (see text).
   $N$ is the number of atoms in the system.
}
\begin{center}
\begin{tabular}{c|cc|cc|}
 & \multicolumn{2}{c|}{$R_c=4$\AA} & \multicolumn{2}{c|}{$R_c=5$\AA} \\
$N$ & CG & SCF & CG & SCF \\
\hline
64   & 5.8 & 9.3 & 8.4 & 8.4 \\
512  & 4.9 & 11.4 & 8.8 & 10.1 \\
1000 & 4.3 & 11.5 & 9.9 & 11.5 \\
\hline
\end{tabular}
\end{center}
\label{iterations}
\end{table}

   Figure~\ref{cpu-mem} shows the essentially perfect ${\cal O}(N)$
behaviour of the overall CPU time and memory.
   This is not surprising in view of the completely strict enforcement 
of ${\cal O}(N)$ algorithms everywhere in the code
(except the marginal $N \log N$ factor in the FFT used to solve
Poisson's equation, which represents a very small fraction of
CPU time even for 4000 atoms).

\begin{figure}[htbp]
\includegraphics[width=\columnwidth] {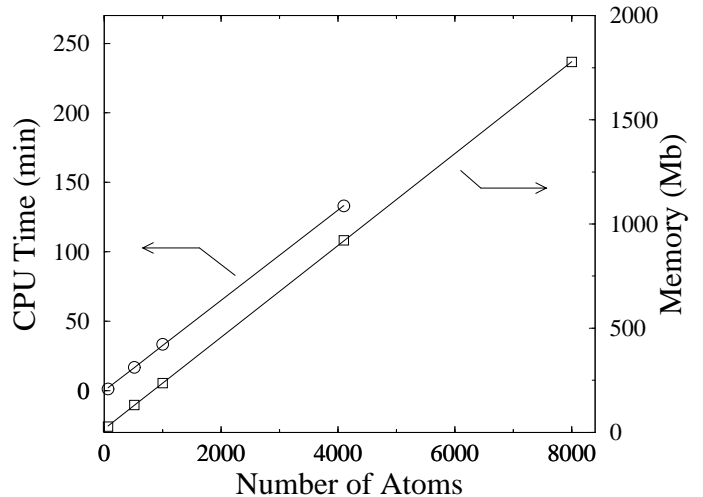}
\caption{   CPU time and memory for silicon supercells of
   64, 512, 1000, 4096, and 8000 atoms.
      Times are for one average molecular dynamics step at 300 K.
      This includes 10 SCF steps, each with 10 conjugate gradient
   minimization steps of the ${\cal O}(N)$ energy functional.
      Memories are peak ones. 
      Although the memory requierement for 8000 atoms was 
   determined accurately, the run could not be performed because 
   of insufficient memory in the PC used.
            }
\label{cpu-mem}
\end{figure}

\section{Other features}

   Here we will simply mention some of the possibilities and features
of the {\sc Siesta} implementation of DFT:
\begin{itemize}
\item
   A general-purpose package
\cite{Garcia-Soler:fdf}, the flexible data format (fdf),
initially developped for the {\sc Siesta} project, allows the
introduction of all the data and precision parameters in a simple 
tag-oriented, order-independent format which accepts different 
physical units.
   The data can then be accessed from anywhere in the program,
using simple subroutine calls in which a default value is specified
for the case in which the data are not present.
   A simple call also allows the read pointer to be positioned in order
to read complex data `blocks' also marked with tags.
\item
   The systematic calculation of atomic forces and stress tensor
allows the simultaneous relaxation of atomic coordinates and
cell shape and size, using a conjugate gradients minimization
or several other minimization/annealing algorithms.
\item
   It is possible to perform a variety of molecular dynamics
simulations, at constant energy or temperature, and at constant
volume or pressure, also including Parrinello-Rahman dynamics
with variable cell shape
\cite{Allen-Tildesley1987}.
   The geometry relaxation may be restricted, to impose
certain positions or coordinates, or more complex constraints.
\item
   The auxiliary program {\sc Vibra} processes systematically 
the atomic forces for sets of displaced atomic positions, 
and from them computes the Hessian matrix
and the phonon spectrum. 
   An interface to the {\sc Phonon} program \cite{Parlinski2001} 
is also provided within {\sc Siesta}.
\item
   A linear response program ({\sc Linres}) to calculate phonon 
frequencies has also been developed
\cite{Pruneda2001}.
   The code reads the SCF solution obtained by Siesta, and 
calculates the linear response to the atomic displacements, 
using first order perturbation theory.
   It then calculates the dynamical matrix, from which the phonon 
frequencies are obtained.
\item
   A number of auxiliary programs allows various representations 
of the total density, the total and local density of states, and the
electrostatic or total potentials. 
   The representations include both two-dimensional cuts
and three-dimensional views, which may be colored to 
simultaneously represent the density and potential.
\item
   Thanks to an interface with the {\sc Transiesta} program,
it is posible to calculate transport properties across a 
nanocontact, finding self-consistently the effective
potential across a finite voltage drop, at a DFT level,
using the Keldish Green's function formalism
\cite{Brandbyge2001}.
\item
   The optical response can be studied with 
{\sc siesta} using different approaches. 
   An approximate dielectric function can be calculated from
the dipolar transition matrix elements between
occupied and unoccupied single-electron eigenstates using
first order time-dependent pertubation theory.\cite{Economou1983}
   For finite systems, these are easily calculated from the matrix 
elements of the position operator between the basis orbitals.
   For infinite periodic systems,  we use the matrix elements of the
momentum operator. 
   It is important to notice, however, that the use of non-local 
pseudopotentials requires some correction terms
\cite{Read-Needs1991}.

   We have also implemented a more sophisticated approach to 
compute the optical response of finite systems, using the adiabatic 
approximation to time-dependent DFT\cite{Gross1995,Gross1996}.
   The idea is to integrate the time-dependent Schr\"odinger
equation when a time depedent perturbation is applied to the
system\cite{Yabana1996}.
   From the time evolution, it is then possible to extract the optical
adsorption and dipole strength functions, including some genuinely
many-body effects, like plasmons.  
   Using this approach we have succesfully calculated the electronic 
response of systems such as fullerenes and small metallic
clusters\cite{Tsolakidis2001}.
\end{itemize}

\section*{ACKNOWLEDGMENTS}

   We are deeply indebted to Otto Sankey and David Drabold for
allowing us to use their code as an initial seed for this project,
and to Richard Martin for continuous ideas and support.
   We thank Jose Luis Martins for numerous discussions
and ideas,  and J\"urgen K\"ubler for helping us implement the 
noncollinear spin.
   The exchange-correlation methods and routines were developed
in collaboration with Carlos Balbas and Jose L. Martins.
   We also thank  In-Ho Lee, Maider Machado, Juana Moreno,
and Art R. Williams  for some routines, and Eduardo Anglada
and Oscar Paz for their computational help.
   This work was supported by the Fundaci\'on Ram\'on Areces 
and by Spain's MCyT grant BFM2000-1312. 
   JDG would like to thank the Royal Society for a University 
Research Fellowship and EPSRC for the provision of computer 
facilities.
   DSP acknowledges support from the Basque Government 
(Programa de Formaci\'on de Investigadores).

\appendix

\section{Orthogonality force and stress}
\label{Orthogonality_force}

   We have yet to comment on the force and stress terms containing 
$\partial\rho_{\mu\nu}/\partial{\bf R}_I$.
   Substituting the first term of Eq.~(\ref{drhodr}) into 
Eqs.~(\ref{dENAdr}-\ref{dExcdr}) and adding the first term of
Eq.~(\ref{dEkdr}) we obtain a simple expression:
$\sum_{\mu\nu} H_{\nu\mu} \partial\rho_{\mu\nu}/\partial{\bf R}_I$.
   Now, $\rho_{\mu\nu}$ is a function of the Hamiltonian eigenvector
coefficients and occupations only (Eq.~(\ref{DM})).
   On the Born-Oppenheimer surface (BOS), $E^{KS}$
is stationary with respect to those coefficients and 
occupations, and the Hellman-Feynmann theorem guarantees that any 
change of them will not modify the total energy to first order, and
therefore will not affect the forces.
   In other words, the atomic forces are the partial derivatives
$\partial E^{KS} / \partial {\bf R}_I$ {\em at constant $c_{\mu i}$
and $n_i$}.
   Even in the Car-Parrinello scheme, in which the system moves
out of the BOS, making the Hellman-Feynmann
theorem invalid, the atomic forces are nevertheless {\em defined}
as derivatives at constant $c_{\mu i}$ and $n_i$.
   Thus, it may seem that the terms
$\partial\rho_{\mu\nu}/\partial{\bf R}_I$ are irrelevant for the
calculation of the forces.
   However, in the previous discussion we have omitted to say 
that the KS energy must be minimized under the constraint of 
orthonormality of the occupied states and that, therefore,
at the BOS the energy is stationary {\em only} with respect 
to changes of $\psi_i$  which do not violate the orthonormality.
   With an atomic basis set, the displacement of atoms (and the
deformation of the unit cell) modifies the basis, and therefore
the occupied states $\psi_i = \sum_\mu \phi_\mu c_{\mu i} $,
even at constant $c_{\mu i}$'s.
   And the change of the states affects their orthonormality.
   Thus, in order to calculate the new total energy, we need
to re-orthonormalize the occupied states, by changing their
coefficients $c_{\mu i}$.
   Schematically, we must solve
\begin{equation}
\label{reorthog}
        \langle \psi_i | \delta S | \psi_j \rangle 
     +  \langle \delta \psi_i | S | \psi_j \rangle
     +  \langle \psi_i | S | \delta \psi_j \rangle = 0
\end{equation}
where $\delta S$ represents the change of $S_{\mu\nu}$ due to
the atomic displacements, and $\delta \psi_i$ the modification 
of $\psi_i$ due to the change of $c_{\mu i}$.
   Without lack of generality, we can expand
$\delta \psi_i$ in the basis of the eigenvectors $\psi_i$
as $\delta \psi_i = \sum_j \psi_j \lambda_{ji}$.
   Substituting this expansion into (\ref{reorthog}) and
using that $\langle \psi_i | S | \psi_j \rangle = \delta_{ij}$
we obtain
$\lambda_{ji} = - \frac{1}{2}
\langle \psi_j | \delta S | \psi_i \rangle$.
   Thus
\begin{equation}
\label{delta_psi}
   | \delta \psi_i \rangle = - \frac{1}{2} \sum_j
   | \psi_j \rangle \langle \psi_j | \delta S | \psi_i \rangle.
\end{equation}
   In terms of the coefficients $c_{\mu i}$, we have
$\langle \psi_j | \delta S | \psi_i \rangle =
\sum_{\mu\nu}  c_{j \mu} \delta S_{\mu\nu} c_{\nu i}$ and
\begin{eqnarray}
\label{delta_cimu}
   \delta c_{\mu i} 
   & = & - \frac{1}{2} \sum_j \sum_{\eta\nu}
            c_{\mu j} c_{j \eta} \delta S_{\eta\nu} c_{\nu i}
            \nonumber \\
   & = & - \frac{1}{2} \sum_{\eta\nu}
             S^{-1}_{\mu\eta} \delta S_{\eta\nu} c_{\nu i}
\end{eqnarray}
where we have used that
$c_{\mu i} = \langle \tilde{\phi}_\mu | \psi_i \rangle$, and
\begin{equation}
   \sum_i c_{\mu i} c_{i \nu} = 
    \langle \tilde{\phi}_\mu | \tilde{\phi}_\nu \rangle = 
   S^{-1}_{\mu\nu}
\end{equation}
   Differentiating now Eq.~(\ref{DM}) and using (\ref{delta_cimu}) 
we obtain
\begin{equation}
\label{delta_DM}
   \delta \rho_{\mu\nu} = - \frac{1}{2} \sum_{\eta\zeta}
      \left( \rho_{\mu\eta} \delta S_{\eta\zeta} S^{-1}_{\zeta\nu}
           + S^{-1}_{\mu\eta} \delta S_{\eta\zeta} \rho_{\zeta\nu} 
      \right)
\end{equation}
   And 
\begin{equation}
\label{dEorthog}
   \sum_{\mu\nu} H_{\mu\nu} \delta \rho_{\nu\mu} =
      - \sum_{\mu\nu} E_{\mu\nu} \delta S_{\nu\mu}
\end{equation}
where $E_{\mu\nu}$ is the so-called energy-density matrix:
\begin{eqnarray}
\label{E_munu}
   E_{\mu\nu} 
     & = & \frac{1}{2} \sum_{\eta\zeta}
              \left( S^{-1}_{\mu\eta} H_{\eta\zeta} \rho_{\zeta\nu}
             + \rho_{\mu\eta} H_{\eta\zeta} S^{-1}_{\zeta\nu}
              \right)
             \nonumber \\
    &  = & \sum_i c_{\mu i}  n_i \epsilon_i c_{i \nu}
\end{eqnarray}
where $\epsilon_i$ are the eigenstate energies.
   To calculate the orthogonalization force or stress, 
$\delta S_{\mu\nu}$ must be substituted by the appropriate derivative:
\begin{equation}
\label{F_orthog}
   {\bf F}_I^{orthog} = 2 \sum_\mu \sum_{\nu \in I} E_{\nu\mu}
      \frac{ \partial S_{\mu\nu} }{ \partial {\bf R}_{\mu\nu} }
\end{equation}
   This equation has been derived before in different ways, 
and Ordej\'on et al \cite{Ordejon1995} found it also for the  
${\cal O}(N)$ functional, even though it does not require the occupied 
states to be orthogonal.
   In this case, Eq.~(\ref{E_munu}) must be substituted by
a more complicated expression
\cite{Ordejon1995}.

   Similarly, the stress contribution is
\begin{equation}
\label{Stress_orthog}
   \sigma_{\alpha\beta}^{orthog} = - \sum_{\mu\nu} E_{\nu\mu}
      \frac{ \partial S_{\mu\nu} }{ \partial R^\alpha_{\mu\nu} }
      R^\beta_{\mu\nu}
\end{equation}

\section{Radial fast Fourier transform}
\label{RadFFT}

   We consider here how to perform fast integrals of the form
\begin{equation}
\label{rfft_def}
   \psi_l(k) = \int_0^{\infty} r^2 dr j_l(kr) \psi_l(r)
\end{equation}
where $j_l(kr)$ is a spherical Bessel function and $\psi_l(r)$ is a
radial function which behaves as
$\psi_l(r) \sim r^l$ for $r \rightarrow 0$.
   Although methods to perform fast Bessel and H\"ankel transforms
have been described previously in different fields
 \cite{Suter1991,Mohsen-Hashish1994,Ferrari1999}
we have developed a simple method adapted to our needs.
   It is based on the fact that $j_l(x)$ has the general form
$(P^s_l(x)\sin(x)+P^c_l(x)\cos(x))/x^{l+1}$, where $P^s_l(x), P^c_l(x)$
are simple polynomials
$ P^{s,c}_l(x) = \sum_{n=0}^l c^{s,c}_{ln} x^n$.
   Thus, the method involves computing $l+1$ fast sine and cosine 
transforms \cite{NumericalRecipes} and add the different terms:
\begin{eqnarray}
\label{rfft_exp}
   \psi_l(k) 
      & = & \sum_{n=0}^l \frac{c^s_{ln}}{2 k^{l+1-n}}
            \int_{-\infty}^{+\infty} \frac{\psi_l(r)}{r^{l-1-n}} \sin(kr) dr 
            \nonumber \\
      & + & \sum_{n=0}^l \frac{c^c_{ln}}{2 k^{l+1-n}}
            \int_{-\infty}^{+\infty} \frac{\psi_l(r)}{r^{l-1-n}} \cos(kr) dr 
\end{eqnarray}
   Notice that we have extended the integral to the whole real axis,
defining $\psi_l(-r) \equiv (-1)^l \psi_l(r)$, in accordance
to the  behaviour $\psi_l(r) \sim r^l, r \rightarrow 0$.
   The coefficients $c^{s,c}_{ln}$ can be obtained by defining a complex
polynomial $P_l(x) = P^c_l(x) + i P^s_l(x)$, which obeys the recurrence
relations
 \cite{Abramowitz-Stegun}
\begin{eqnarray}
\label{P}
   && P_0(x) = i \equiv \sqrt{-1} \nonumber \\
   && P_1(x) = i - x              \\
   && P_{l+1}(x) = (2l+1) P_l(x) - x^2 P_{l-1}(x)  \nonumber
\end{eqnarray}
   In order to perform the integrals in (\ref{rfft_exp}) using discrete
FFT's, we need to calculate $\psi(r)$ on a regular radial grid, up to
a maximum radius $r_{max}$, beyond which $\psi(r)$ is assumed to be 
strictly zero.
   The separation $\Delta r$ between grid points determines
a cutoff $k_{max} = \pi / \Delta r$ in reciprocal space, and vice versa,
$\Delta k = \pi / r_{max}$.
   For convolutions, such as those involved in Eq.~(\ref{Sl1l2l}),
we need $r_{max} = r^c_1 + r^c_2$ and $k_{max} = \max(k^c_1,k^c_2)$,
where $r^c_{1,2}, k^c_{1,2}$ are the cutoff radii and maximum 
wavevectors of $\psi_{1,2}$, respectively. 
   We must then pad with zeros the intervals $[ r^c_{1,2}, r_{max} ]$
for the forward transforms $\psi_{1,2}(r) \rightarrow \psi_{1,2}(k)$.
   In practice, we set 
$r_{max} = 2 \max_{\mu}(r^c_{\mu}), k_{max} = \max_{\mu}(k^c_{\mu})$, 
where $\mu$ labels all the basis orbitals and KB projectors, 
and we use the same real and reciprocal grids for all orbital pairs.
   In this way, we need to perform the forward transform only
once for each radial function $\psi_{\mu}(r)$.
   Finally, notice that in Eq.~(\ref{Sl1l2l}), 
$\psi^*_{1,l_1 m_1}(k) \psi_{2,l_2 m_2}(k) \sim k^{l_1+l_2}$ for
$k \rightarrow 0$, while $l_1 + l_2 -l$ is even and nonnegative, so that 
the integrands of Eq.~(\ref{rfft_exp}) for the backward transform are 
all even and well behaved at the origin.

\section{Extended-mesh algorithm}

   We describe here a simple and efficient algorithm to handle mesh
indices in three-dimensional periodic systems.
   Its versatility makes it suitable for several different tasks in {\sc Siesta}
like neighbor-list constructions, basis orbital evaluation in the
real-space integration grid, density-gradient calculations in the GGA, 
etc.
   It would be also very apropriate for other problems, like the solution of 
partial differential equations by real-space discretization or the 
calculation of the interaction energy in lattice models.
   For clarity of the exposition, we will describe the algorithm for a
specially simple application, namely the evaluation of the
Laplacian of a function $f({\bf r})$ using finite differences, even though 
the algorithm is not used in {\sc Siesta} for this purpose.
   In three dimensions, one generally discretizes space in
all three periodic directions, using an index for each direction.
   For simplicity, let us consider an orthorhombic unit cell, with 
mesh steps $\Delta x, \Delta y, \Delta z$.
   Then the simplest formula for the Laplacian is
\begin{eqnarray}
  \nabla^2 f_{i_x,i_y,i_z} &=& 
      (f_{i_x+1,i_y,i_z} - 2 f_{i_x,i_y,i_z} + f_{i_x-1,i_y,i_z}) / \Delta x^2
     \nonumber \\
  &+& (f_{i_x,i_y+1,i_z} - 2 f_{i_x,i_y,i_z} + f_{i_x,i_y-1,i_z}) / \Delta y^2
     \nonumber \\
  &+& (f_{i_x,i_y,i_z+1} - 2 f_{i_x,i_y,i_z} + f_{i_x,i_y,i_z-1}) / \Delta z^2
     \nonumber
\end{eqnarray}
   A direct translation of this expression into Fortran90 code might read
\begin{verbatim}
   Lf(ix,iy,iz) =                         &
        ( f(modulo(ix+1,nx),iy,iz) +      &
          f(modulo(ix-1,nx),iy,iz) )/dx2  &
      + ( f(ix,modulo(iy+1,ny),iz) +      &
          f(ix,modulo(iy-1,ny),iz) )/dy2  &
      + ( f(ix,iy,modulo(iz+1,nz)) +      &
          f(ix,iy,modulo(iz-1,nz)) )/dz2  &
      - f(ix,iy,iz) * (2/dx2+2/dy2+2/dz2)
\end{verbatim}
where the indices $i_\alpha$ ($\alpha=\{x,y,z\}$) of the arrays 
{\tt f} and {\tt Lf} run from 0 to $n_\alpha-1$, as in C.
   There are two problems with this construction. 
   First, the {\tt modulo} operations are required to bring the indices back 
to the allowed range $[0,n_\alpha-1]$.
   And second, the use of three indices to refer to a mesh point implies
implicit arithmetic operations, generated by the compiler, to translate
them into a single index giving its position in memory.

   A straightforward solution to these inefficiencies would be to create
a neighbor-point list \verb'j_neighb(i,neighb)', of the size of the number
of mesh points times the number of neighbor points.
   However, although the latter are only six in our simple example, they 
may frequently be as many as several hundred, what generally makes this
approach unfeasible.
   A partial solution, addressing only the first problem, is to create six
(or more for longer ranges) one-dimensional tables
$j_\alpha^{\pm 1}(i_\alpha) = \mbox{mod}(i_\alpha \pm 1,n_\alpha)$
to avoid the modulo computations~\cite{Binder1992}.
   Here, we describe a multidimensional generalization of this method, 
which solves both problems at the expense of a very reasonable amount 
of extra storage.

   The method is based on an {\em extended mesh}, which extends
beyond the periodic unit cell, by as much as required to cover all the space
that can be reached from the unit cell by the range of the interactions
or the finite-difference operator.
   The extended mesh range is
$i_\alpha^{min}=-\Delta n_\alpha$ and 
$i_\alpha^{max}=n_\alpha - 1 + \Delta n_\alpha$, where 
$\Delta n_\alpha = 1$ in our particular example, in which the Laplacian 
formula extends just to first-neighbor mesh points.
   In principle, in cases with a small unit cell and a long range, the mesh 
extension may be larger than the unit cell itself, extending over several
neighbor cells.
   However, in the more relevant case of a large system, we will expect
the extension region to be small compared to the unit cell.
   We then consider two combined indices, one associated to the normal
unit-cell mesh, and another one associated to the extended mesh
$$
  i = i_x + n_x i_y + n_x n_y i_z, 
$$
$$
  i_{ext} = (i_x-i_x^{min}) + n_x^{ext} (i_y-i_y^{min})
           + n_x^{ext} n_y^{ext} (i_z-i_z^{min}), 
$$
where 
$n_\alpha^{ext}=i_\alpha^{max}-i_\alpha^{min}+1=n_\alpha+2\Delta n_\alpha$.
   The key observation is that, if $i_{ext}$ is a mesh point {\em within}
the unit cell ($0 \le i_\alpha \le n_\alpha-1$), 
and if $j_{ext}$ is a neighbor mesh point (within its interaction range, 
i.e. $| j_\alpha - i_\alpha | \le \Delta n_\alpha$), then the 
arithmetic difference $j_{ext}-i_{ext}$ depends only on the relative
positions of $i_{ext}$ and $j_{ext}$ (i.e. on $j_\alpha-i_\alpha$), 
but not on the position of $i_{ext}$ within the unit cell.
   We can then create a list of neighbor strides $\Delta ij_{ext}$, 
and two arrays to translate back and forth between $i$ and $i_{ext}$.
   One of the arrays maps the unit cell points to the central region of
the extended mesh, while the other one folds back the extended mesh points 
to their periodically equivalent points within the unit cell.
   Then, to access the neighbors of a point $i$, we 
$a$) translate $i \rightarrow i_{ext}$;
$b$) find $j_{ext} = i_{ext} + \Delta ij_{ext}$; and
$c$) translate $j_{ext} \rightarrow j$.
   Notice that several points of the extended mesh will map to the same point
within the unit cell and that, in principle, a unit cell point $j$ may be 
neighbor of $i$ through different values of $j_{ext}$.
   In our example, the innermost loop would then read
\begin{verbatim}
 Lf(i) = 0
 do neighb = 1,n_neighb
    j_ext = i_extended(i) + ij_delta(neighb)
    j = i_cell(j_ext)
    Lf(i) = Lf(i) + L(neighb) * f(j)
 end do
\end{verbatim}
where the number of neighbor points would be \verb'n_neighb=7', 
including the central point itself, and
\begin{verbatim}
 ij_delta(1) =  1             ; L(1) = 1/dx2
 ij_delta(2) = -1             ; L(1) = 1/dx2
 ij_delta(3) =  nx_ext        ; L(3) = 1/dy2
 ij_delta(4) = -nx_ext        ; L(4) = 1/dy2
 ij_delta(5) =  nx_ext*ny_ext ; L(5) = 1/dz2
 ij_delta(6) = -nx_ext*ny_ext ; L(6) = 1/dz2
 ij_delta(7) =  0  ; L(7)=-2/dx2-2/dy2-2/dz2 
\end{verbatim}
   Notice that the above loop is completely general, for any
linear operator, using an arbitrary number of neighbor points for its
finite difference representation.
   In fact, it is even independent of the space dimensionality.
   Furthermore, the index operations required are just one 
addition and three memory calls to arrays of 
range one~\cite{other_algorithms}.
   This inner loop simplicity comes at the expense of the two extra arrays 
\verb'i_extended' and \verb'i_cell' (of the size of the normal and extended 
meshes, respectively) which are generally an acceptable memory overhead.
   Notice, however, that the the neighbor-point list \verb'ij_delta'
is independent of the mesh index $i$, what makes this array quite small
in most problems of interest.

\section{Sparse matrix techniques}

   We will describe here some of the sparse-matrix multiplication
techniques used in evaluating Eqs.~(\ref{DM2rho}), 
(\ref{E_KMG}), (\ref{rho_KMG}), and (\ref{gradE_KMG}).
   There is a large variety of sparse-matrix representations and
algorithms, each one optimized for a different kind of sparsity.
   The main constraint for choosing our representation and 
algorithms is that they must be ${\cal O}(N)$ in both 
memory and CPU time.
   We enforce this condition strictly by requiring, for example,
that a vector of size $\sim N$ will not be reset to zero a number
$\sim N$ of times.
   In our sparse matrices, like $S_{\mu \nu}$, $H_{\mu \nu}$, 
$c_{\mu i}$, $\rho_{\mu \nu}$, and $\phi_\mu({\bf r})$, the number
$p$ of non-zero elements in a row is typically much larger than one
(but still of order $\sim N^0$) and much smaller than the row size 
$m \sim N¹^1$.
   Such matrix rows are efficiently stored as a real vector
of size $p$, containing the non-zero elements, 
and an integer vector of the same size containing
the column index of each non-zero element.
   The whole matrix $A$ of $n$ rows is then represented by two
arrays \verb'A' and  \verb'jcol', of size $n \times p$, such that 
\cite{fortran_order}
$A_{ij}=$~\verb'A(i,k)', where $j=$~\verb'jcol(i,k)'.
   The problem with this representation is that, given a value $j$
of the column index, there is no simple way to access the 
element $A_{ij}$ without scanning the whole row, what is
frequently too costly.
   One solution is to unpack a row $i$, that will be repeatedly used,
into `expanded form', i.e. to transfer it to a vector \verb'Arow'
of the full row size $m$ (containing also all the zeros),
so that $A_{ij}=$~\verb'Arow(j)'.
   Since  $p>>1$, the size of \verb'Arow' is negligible compared 
to that of  \verb'A' and  \verb'jcol'.

   To find the matrix product $C$ of  two sparse matrices $A$ and $B$ 
$$
  C_{ik} = \sum_j A_{ij} B_{jk}
$$
we proceed iteratively for each row $i$ of $A$
(which will generate the same row of $C$): each non-zero
element $j$ of the row is multiplied by every non-zero element
of the $j$th row of $B$ (whose column index is, say $k$) and the 
result is accumulated in the $k$'th position of an auxiliary `expanded' 
vector.
   After finishing with that row of $A$ we pack the vector in
sparse format into the $i$th row of $C$ and restore the 
auxiliary vector to zero.
   In fact, the packing can be performed simultaneously with
the product, using an auxiliary index vector instead:
\begin{verbatim}
 C = 0.
 ncolC = 0
 jcolC = 0
 pos = 0  ! Auxiliary index vector
 do i = 1, nA
   do jA = 1,ncolA(i)
     j = jcolA(i,jA)
     do jB = 1,ncolB(j)
       k = jcolB(j,jB)
       jC = pos(k) 
       if (jC==0) then ! New non-zero col
         ncolC(i) = ncolC(i) + 1
         jC = ncolC(i)
         jcolC(i,jC) = k
         pos(k) = jC
       endif
       C(i,jC) = C(i,jC) + A(i,jA)*B(j,jB)
     enddo
   enddo
   do jA = 1,ncolA(i) ! Restore pos to zero
     j = jcolA(i,jA)
     do jB = 1,ncolB(j)
       k = jcolB(j,jB)
       pos(k) = 0
     enddo
   enddo
 enddo
\end{verbatim}
   Notice that the auxiliary vector \verb'pos', which keeps the
position in `packed format' of the non-zero elements of
one row of $C$, is initiallized in full only once.
   Notice also that this algorithm, unlike those of
ref.~\onlinecite{NumericalRecipes}, does not require the
matrix elements to be stored in ascending column order.

   The previous algorithm generates all the non-zero elements
of  $C$ but in many cases we need only some of them.
   For example, to calculate the electron density 
(Eq.~(\ref{DM2rho})), we need only the density matrix elements 
$\rho_{\mu \nu}$ for which $\phi_\mu$ and $\phi_\nu$ overlap.
   Also the expression (\ref{gradE_KMG}) needs to be evaluated
only for the coefficients $c_{\mu i}$ which are allowed to be
non-zero by the localization constraints.
   In these cases, in which the array \verb'jcolC' is already 
known, another algorithm is more effective.
   We start by finding the sparse representation of $B$ in 
{\em column} order or, in other words, the transpose of  $B$:
\begin{verbatim}
 Bt = 0 ! B transpose
 jcolBt = 0
 ncolBt = 0
 do i = 1,nB
   do jB = 1,ncolB(i)
     j = jcolB(i,jB)
     ncolBt(j) = ncolBt(j) + 1
     jBt = ncolBt(j)
     jcolBt(j,jBt) = i
     Bt(j,jBt) = B(i,jB)
   enddo
 enddo
\end{verbatim}
   We then unpack a row $i$ of $A$ and multiply it by a column 
$j$ of $B$ (a row of its transpose) for each required matrix element 
$C_{ij}$ of their product:
\begin{verbatim}
 C = 0.
 Arow = 0.  ! Auxiliary vector
 do i = 1,nC
   do jA = 1,ncolA(i) ! Copy one row of A
     j = jcolA(i,jA)
     Arow(j) = A(i,jA)
   enddo
   do jC = 1,ncolC(i) ! Calculate Cij
     j = jcolC(i,jC)
     do jBt = 1,ncolBt(j)
       k = jcolBt(j,jBt)
       C(i,jC) = C(i,jC) + Arow(k)*Bt(j,jBt)
     enddo
   enddo
   do jA = 1,ncolA(i) ! Restore Arow to zero
     j = jcolA(i,jA)
     Arow(j) = 0.
   enddo
 enddo
\end{verbatim}
   The combination of these two matrix multiplication algorithms
allows an efficient evaluation of 
Eqs.~(\ref{E_KMG}-\ref{gradE_KMG}).
   Since these equations involve a trace or a relatively small
subset of a final matrix, it is important to control the order 
and sparsity of the intermediate products, in order to keep
them as sparse as possible.
   Notice that, once a row of $A \times B$  has been evaluated, 
it may be multiplied by a third matrix, to obtain a row of the final 
product, without need of storing the whole intermediate matrix.

   To calculate the density at a grid point using Eq.~(\ref{DM2rho})
we need to access the matrix elements $\rho_{\mu \nu}$, and
this is inefficient if they are stored in sparse format.
   Thus, we first copy the matrix elements, between the $n_r$
basis orbitals which are non-zero at the grid point $r$, into an
auxiliary matrix array, of size $n_{aux} \times n_{aux}$, 
with $n_{aux} \ge n_r$.
   We also create a lookup table \verb'pos', of size equal to the 
total number of basis orbitals, such that \verb'pos(mu)' is the
position, in the auxiliary matrix, of the matrix elements of orbital
\verb'mu' (or zero if they have not been copied to it).
   If there are new non-zero orbitals at the next grid points, we
keep copying them into the auxiliary matrix, until all its $n_{aux}$
slots are full, at which point we erase it and restart the process.
   Since succesive grid points tend to contain the same non-zero
basis orbitals, these copies and erasures are not frequent.

\bibliographystyle{apsrev}
\bibliography{siesta_paper}

\begin{thebibliography}{86}
\bibitem[{Vlo({\natexlab{b}})}]{Vlocal2}
\bibinfo{note}{For some atoms, tipically those with semicore states suitable to
  be treated together with the valence states, $V_l(r)$ only assumes the
  asymptotic coulombic behaviour $-2Z_{val} / r$, and therefore only cancels
  out exactly with our $V_{local}(r)$, for $r$ larger than certain $r_C >
  r_{core}$. In these cases, to avoid very extended KB projector functions, we
  generate the local potentials with a prescription different from that
  presented in the text: if $r_C > 1.3~r_{core}$ we take, $V_{local}(r)=V_l(r)$
  for $r> r_{core}$ and, $V_{local}(r)=\exp(v_1+v_2r^2+v_3r^3)$ for $ r <
  r_{core}$, where $v_1$, $v_2$, and $v_3$ are determined by enforcing the
  continuity of the potential up to the second derivative. This simple
  prescription usually produces smooth local potentials with similar properties
  to those commented in the text (see note \onlinecite{Vlocal1})}.
\bibitem[{ion()}]{ionicPS}
\bibinfo{note}{If there are both semicore and valence electrons with the same
  angular momentum, the pseudopotential is generated for an ion.}
\bibitem[{den()}]{densityEcut}
\bibinfo{note}{Notice that our grid cutoff to represent the density is not
  directly comparable to the energy cutoff in the context of plane wave codes,
  which usually refers to the wavefunctions. Strictly speaking, the density
  requires a value four times larger.}
\bibitem[{oth()}]{other_algorithms}
\bibinfo{note}{In the present example, further savings can be achieved by
  extending the arrays f(i) and Lf(i) themselves, eliminating the index
  translations and facilitating the paralelization. Another efficient
  possibility is provided in Fortran90 by the intrinsic cshift operation.
  However, these approaches are more complicated in other cases, like
  atomic-neighbor list constructions, while the double index algorithm is quite
  general.}
\bibitem[{for()}]{fortran_order}
\bibinfo{note}{In Fortran, we alternate the order of $i$ and $k$, to store the
  row elements consecutively in memory. If the number of nonzero elements
  fluctuates largely for different rows, we also store all the rows
  consecutively as a large single vector.}
\end{thebibliography}

\end{document}